\documentstyle[11pt,aaspp4,natbib]{article}

\def\etal{et al.}
\def\eg{e.g.}
\def\gadget{{\small GADGET}}
\def\scf{{\small SCF}}
\def\amax{$\alpha_{max}$}
\def\omh{$\Omega_h$}

\def\aave{$\langle\alpha\rangle$}
\def\ie{{\it i.e.}}
\def\kms{km s$^{-1}$}
\def\kmskpc{km s$^{-1}$ kpc$^{-1}$}
\def\blah2max{$\delta_2^{max}$}
\def\d2ave{$\langle \delta_2^{max}\rangle$}
\def\pave{$\langle p \rangle$}
\def\aavet{$\langle\alpha\rangle_t$}
\def\vrot{$v_{rot}$}
\def\mratio{$(M_d/M_h)_{R_\odot}$}
\def\ep{${\mathcal E}^\prime$}

\begin{document}

\title{The Influence of Triaxial Halos on Collisionless Galactic Disks}
\author{Jeremy L. Tinker and Barbara S. Ryden}
\affil{Department of Astronomy, Ohio State University,\\ 
	140 W. 18$^{th}$ Avenue, Columbus, OH 43210, USA}
\affil{E-mail: tinker, ryden@astronomy.ohio-state.edu}

\begin{abstract}

We investigate the effect of rotating, triaxial halos on disk galaxies
through an extensive set of numerical $N$-body simulations. Our
simulations use a rigid potential field for the halos and bulges and
collisionless particles for the disks. The triaxiality and the
rotation rate of the halo are varied, as well as the masses of all
three galaxy components. We analyze both the bar stability and the
spiral response of the disks under these conditions.

We characterize most of our models by the mass ratio of the disk to
the halo at 2.3 disk scale lengths, \mratio. For models with a mass
ratio greater than 0.8, a halo pattern speed \omh\ = 6.7 \kmskpc, and
a intermediate-to-major axis ratio $q_b=0.85$, a strong bar will
develop within 3 Gyr, even for models with a bulge mass
$M_b=0.3M_d$. Models in which the bulge mass is reduced by half
develop bars earlier and with lower \omh.

We create an artificial Hubble sequence of disk galaxies by varying
the bulge-to-disk ratio of our models from 0 to 2.5. The torque
induced by a rotating, non-axisymmetric halo creates bisymmetric
spiral structure in the disk.  We find that the pitch angle of the
spiral arms in these models follows the same general trend found in
observations of spiral galaxies, namely that later type galaxies have
higher pitch angles. Our simulations follow closely the
observational relation of spiral pitch angle with maximum rotational
velocity of the disk, where galaxies with faster rotation have more
tightly wound spiral arms. This relation is followed in our
simulations regardless of whether the dominating mass component of the
galaxy is the disk, the halo, or the bulge.


\end{abstract}

\section{Introduction}

Galaxies are not spherical objects. Although it is difficult to
measure the intrinsic shapes of astronomical bodies from their
projected light distributions, inferences can be make from their
observed ellipticities. Many studies of the distribution of observed
axis ratios of galaxies and galaxy clusters (\eg\ Ryden, 1992, 1996;
Tremblay \& Merritt 1996) have shown that they are well represented by
a population of intrinsically triaxial objects. Similar statistical
analysis performed by Alam \& Ryden (2002) with a much larger sample
of galaxies from the Sloan Digital Sky Survey Early Data Release has
strengthened the claim that the observed distributions of
ellipticities cannot be fit by projections of oblate or prolate
spheroids.

This type of analysis is only available in the cases where the mass
distribution creates light. The shapes of dark matter halos,
therefore, must be inferred from even more indirect means. Proxy
measures of the shape of our own halo offer widely varying
results. The tidal stream of the Sagittarius dwarf galaxy implies a
nearly spherical halo (Ibata \etal\ 2001), in contrast to the results
of star counts, which give a minor to major axis ratio of $\sim$ 0.6
(Siegal \etal\ 2002). The dark matter halos created in numerical
studies of structure formation are non-axisymmetric systems. $N$-body
simulations of dissipationless collapse in the cold dark matter (CDM)
scenario by Dubinski \& Carlberg (1991) showed that the resulting
virialized halos deviated substantially from spherical symmetry and
from axisymmetry, with average axis ratios $\langle c/a\rangle=0.5$
and $\langle b/a\rangle=0.7$, where $a$, $b$ and $c$ are the major,
intermediate, and minor axes respectively. Simulations of larger
cosmological volumes, which more adequately reproduce the hierarchical
nature of structure formation in CDM, have also shown that dark matter
halos are triaxial (Barnes \& Efstathiou 1987; Warren \etal\
1992). Analysis of the currently favored $\Lambda$CDM cosmology has
shown similar results (Jing \& Suto 2002).

The simulations cited above consist only of a collisionless dark
matter component and include no gas dynamics. The introduction of a
gas and stellar disk in the equatorial plane of the halo would
certainly have effects on the internal structure of the halo. Fully
self-consistent cosmological simulations of dissipational structure
formation have problems reproducing more than general galactic
attributes (\eg\ Murali \etal\ 2002), and the resulting halo shape has
not been the focus of these efforts.  However, Dubinski (1994) showed
that the adiabatic growth of a disk-like potential in a triaxial halo
did not influence $\langle c/a\rangle$, while the intermediate to major axis
ratio was only slightly increased to $\langle b/a\rangle \gtrsim 0.7$.

Models and simulations of disk galaxies, however, usually employ
spherical halos, which do not reflect these results from CDM
simulations. The purpose of most current investigations of
galaxy-scale simulations has been modeling transformations of galaxies
through major mergers (\eg\ Barnes 1992; Dubinski, Mihos, \& Hernquist
1999), minor encounters (\eg\ Struck 1997; Thakar \& Ryden 1998) and
gas-dynamical effects (\eg\ Mihos \& Hernquist 1996; Springel
2000). The triaxiality of the halo does not immediately effect the
results of these simulations and the implications of using triaxial
halos have not been explored.

Early studies of the stability of disk galaxies (e.g. Ostriker \&
Peebles 1973; Hohl 1976; Efstathiou \etal\ 1982) demonstrated that a
cold, rotationally supported disk is dramatically unstable to bar
formation. This fact was used as justification for the existence of
heavy, spherical, dark matter halos enveloping disk galaxies. This
dynamically hot component of the galaxy inhibits bar formation. It has
also been shown that a dense core gives a galaxy an inner Lindblad
resonance, which does not allow swing amplification of waves through
the center of the galaxy and thus acts as a bar suppressant (Sellwood
1989)\footnote{Efstathiou \etal\ (1982) did perform two simulations
with dense centers but found that it did not make their disks stable.
Sellwood proposed that the low resolution of their simulations was
responsible for the results of their ``bulge'' simulations.}. In all
cases the stabilizing mass component has been assumed to have
spherical symmetry. If these galaxies were enveloped by triaxial
halos, they would be more susceptible to bar instabilities. A
systematic exploration of the stability of disks with non-axisymmetric
halos has not been conducted. Curir \& Mazzei (1999) presented fully
self-consistent gas-dynamical simulations of disk galaxies with
triaxial halos, but only used two halo models and did not employ
bulges. The resolution in their simulations was severely limited
($N_{disk}=3,000$), far less than the number of particles used by
Efstathiou \etal\ (1982) which Sellwood (1989) found to be
insufficient.

The notion that the potential field of a galaxy may contain
non-axisymmetric features is not a new one; it is, in fact, readily
apparent from the bars seen in many galactic disks. The effect of an
oval distortion, like a bar, on a gas disk has been proposed as a
driving mechanism for spiral density waves (Lin 1970). This idea has
been explored numerically as well (Sanders \& Huntley 1976; Sanders
1977; Huntley \etal\ 1978). These early simulations generally were
limited to two-dimensional, massless disks driven by an analytic
distortion to the symmetric potential, but were successful in creating
significant spiral response in the disk. However, bar forcing does not
reproduce tightly bound spiral patterns. And in comparison to Sanders
\& Huntley (1976), the response of the outer disk is weak when a more
realistic bar potential, one that drops off faster with radius, is
used in the simulation (Sanders \& Tubbs 1980). Observationally spiral
patterns are as common in barred galaxies as galaxies without bars.
Also, there is evidence that bars and spiral arms have different
pattern speeds, showing that one feature may not drive the other
(Sellwood \& Sparke 1988). A triaxial halo would not be subject to
these concerns since the potential of the dark matter would dominate
the outer regions of the disk, especially low mass disks which are
stable to bar formation.

The bulge-to-disk ratio of a spiral galaxy, as well as the tightness
of the spiral pattern, are two integral criteria of the Hubble
classification scheme. On the average, there exists a smooth increase
in pitch angle with later Hubble types (Kennicutt 1981). Such a
correlation, however, is little more than a consistency check between
classification parameters, and there exists significant scatter in the
measured pitch angle for a given Hubble type.

Even with this scatter, the correlation between arm pattern and galaxy
type has prompted several theoretical studies on the origin of the
this relation (see Kennicutt \& Hodge 1982, and references
therein). Kennicutt (1981) found that the correlation between pitch
angle and maximum rotational velocity of the disk was as good as the
theoretical model predictions, suggesting that the mass distribution
and its resulting rotation curve is an important determinant in
producing the shape of spiral arms. A more massive or more concentrated
bulge will induce higher rotation velocities and loosely follow the
correlation with Hubble type as well.

Our simulations include a collisionless stellar disk and a stiff,
rotating halo. Most simulations include a bulge, which is also
stiff. Our use of dissipationless simulations with halos which are not
fully self-consistent allows us to cover a significant range of
parameter space for all three galaxy components; the disk, bulge, and
halo.

The structure of this paper is as follows: \S 2 presents our initial
conditions as well as the $N$-body techniques used in this study. \S 3
shows the analysis and results of our fiducial simulation. \S 4
describes our suite of different galaxy models. In \S 5 we investigate
the stability of our models with different halo and bulge masses, as
well as different halo rotation rates. \S 6 presents our results for
the spiral morphology of our simulations. We present results for
models which follow the bulge-to-disk ratios of the Hubble sequence
and models which vary widely in rotation velocity. We compare our
results to observations of spiral galaxies.

\section{Numerical Methods}

\subsection{Initial Conditions}

The positions and velocities of the disk particles are initialized in
the method outlined by Hernquist (1993). The density distribution of
the disk is described by

\begin{equation}
\rho(R,z)=\frac{M_d}{4\pi h^2z_0}\exp\left(-\frac{R}{h}\right)\,\mbox{sech}^2\left(\frac{z}{z_0}\right)
\end{equation}

\noindent where $R$ and $z$ are cylindrical coordinates, $h$ and $z_0$ are the
scale length and scale height of the disk, and $M_d$ is the total mass
of the disk. The circular velocity and velocity dispersions are
obtained from

\begin{equation}
v_c=\sqrt{|a_R| R}
\end{equation}

\begin{equation}
\sigma_R^2=\sigma_{R,0}^2 \exp(-R/h)
\end{equation}

\begin{equation}
\sigma_z^2=\pi G \Sigma(R)z_0
\end{equation}

\begin{equation}
\sigma_\phi^2=\sigma_R^2\frac{\kappa^2}{4\Omega^2}
\end{equation}

\noindent where $a_R$ is radial acceleration, $\Sigma$ is the surface
density of the disk, $\kappa$ is the epicyclic frequency and $\Omega$
is the angular rotation rate. The radial acceleration in equation (2)
is the total acceleration contributed by the self-gravity of the disk,
the dark matter halo, and a spherical bulge. The normalization of the
radial velocity dispersion, $\sigma_{R,0}^2$, is obtained by requiring
$\sigma_R$ at a specified radius $R_0$ to be a multiple of the
critical dispersion at that radius:

\begin{equation}
\sigma_{R=R_0}=Q \frac{3.36G\Sigma(R_0)}{\kappa(R_0)}
\end{equation}

\noindent where $Q$ is Toomre's parameter for gravitational stability
in a rotationally supported disk. The disk models in this paper are
defined to have $Q=1.5$ at $R=2.43h$, which in the Milky Way galaxy
would be approximately the solar radius. The rotational velocity of
each particle is then given by

\begin{equation}
\overline{v_\phi}^2-v_c^2=\sigma_R^2\left(1-\frac{\kappa^2}{4\Omega^2}-2\frac{R}{h}\right).
\end{equation}

The different halo models used in this paper can be represented by
Dehnen's (1993) $\gamma$-models;

\begin{equation}
\rho(r)=M_h\frac{3-\gamma}{4\pi}\frac{r_0}{r^\gamma(r+r_0)^{4-\gamma}}
\end{equation}

\noindent where $\gamma$ is the interior logarithmic slope of the
density profile and $r_0$ is the scale length of the halo. A model
with $\gamma=1$ corresponds to a model proposed by Hernquist (1990)
for bulges and elliptical galaxies. A spherical Hernquist model is
used here to represent the bulges of our galaxies. Dubinski \&
Carlberg (1991) showed that the Hernquist model is a good fit for CDM
halos, and is close to the general density profile for dark halos
found by Navarro, Frenk, \& White (1997), which also asymptotes to
$\sim r^{-1}$ at small radii. The $\gamma$-models can be generalized
to triaxial systems by replacing the radial coordinate $r$ with $m$,
where

\begin{equation}
m^2=\frac{x^2}{a^2}+\frac{y^2}{b^2}+\frac{z^2}{c^2}
\end{equation}

\noindent where $a$, $b$, and $c$ are once again the principal axes of the
system.

In practice, the disk-particle positions are sampled from the density
function in equation (1), the accelerations on each particle from the
different galaxy components are computed, and then the velocity
dispersions and the rotational velocity of each particle are
calculated. The peculiar velocities of each particle are sampled from
Gaussian probability distributions with dispersions given by equations
(3)-(5). Because $\kappa$ and $\Omega$ in equations (5)-(7) are assumed to
be azimuthally symmetric, the velocities for the disk particles are
computed assuming that the halo is spherical; therefore when the disk
models are placed inside a triaxial halo they are initially out of
equilibrium.

\subsection{N-body Techniques}

To compute the self-gravity of the disk particles, we have employed
\gadget\ (Springel, Yoshida, \& White 2001), a tree code based on the
hierarchical algorithm of Barnes \& Hut (1986). \gadget\ employs a 
spline-softened force calculation that is expanded to quadrupole
order for particle-cell interactions. The timesteps are continuously
varying and individual for each particle. The spatial and time
resolutions are set by the smoothing length, $\epsilon$, and the
timestep accuracy parameter, $\alpha$. For simulations with $N=10^5$,
which are the majority of the simulations in this paper, these
parameters were set to $\epsilon=0.057$ and $\alpha=0.04$.

The gravitational contribution of the halo is computed by the
self-consistent field (\scf) method of Hernquist \& Ostriker
(1992). This method is based on the expansion of the galaxy potential
in an orthonormal set of basis functions. The accuracy of the
potential calculated is driven by the number of terms in the
expansion. The numbers $n$ and $l$ determine the number of radial and
angular basis functions used. The integration time, however, is a
linear function of the number of terms as well. We found that the best
balance of accuracy and speed was achieved with $n$=16 and $l$=10,
which were the values used in all the simulations. The zeroth-order
basis function was chosen to be the Hernquist profile, and the
coefficients of the expansion were computed by Monte Carlo sampling of
the halo density distribution, $\rho(m)$. Since the coefficients only
have to be calculated once for each halo model, accurate calculations
can be made by initially sampling a large number of particles from
each halo profile, $\sim 10^7$ particles. The \scf\ method has been
used frequently to model triaxial systems (\eg\ Merritt \& Quinlan
1998; Holley-Bockelmann \etal\ 2001). When comparing the force on a set
of test particles calculated from the \scf\ coefficients with the
force calculated from direct summation over the particles used for the
coefficients, the rms error was $\sim 6\times10^{-3}$. This error
level was maintained in models where $\gamma$ was not equal to unity.

The contribution to the potential of the galaxy from the bulge can be
computed analytically. Therefore, the total acceleration of each
particle is given by

\begin{equation}
{\bf a}_i={\bf a}_{tree}\,-\,\frac{M_b}{(r+r_b)^2}{\bf \hat{r}}\,-\,\sum_{n,l,m}A_{nlm}\nabla\Phi_{nlm}.
\end{equation}

\noindent The first term is the acceleration returned by \gadget\ from
walking the tree of disk particles. The middle term is the
acceleration from the bulge, with the subscript $b$ denoting bulge
properties. The third term in equation (10) is the acceleration
computed from the \scf\ expansion of the halo potential. The particle
timesteps are inversely proportional to the particle acceleration,
$\Delta t_i=\alpha/|{\bf a}_i|$.

\begin{deluxetable}{ccc}
\tablecolumns{3} 
\tablewidth{22pc} 
\tablecaption{Properties of the Fiducial Simulation} 
\tablehead{ 
\colhead{Disk Properties} & {Halo Properties} & {Bulge Properties}
}
\startdata
$M_d$ = 1.0	& $M_h$ = 6.0 	& $M_b$ = 0.3 \\
$h$ = 1.0	& $R_0$ = 3.63	& $r_b$ = 0.2 \\
$z_0$ = 0.2	& $\gamma$ = 1.0	& $\gamma$ = 1.0 \\
		& $q_b$ = 0.85 & \\
		& $q_c$ = 0.70 & \\
		& $\Omega_h$ = 0.06 & \\

\enddata
\end{deluxetable}

\begin{figure}[t]
\vspace{8.5cm}
\includegraphics{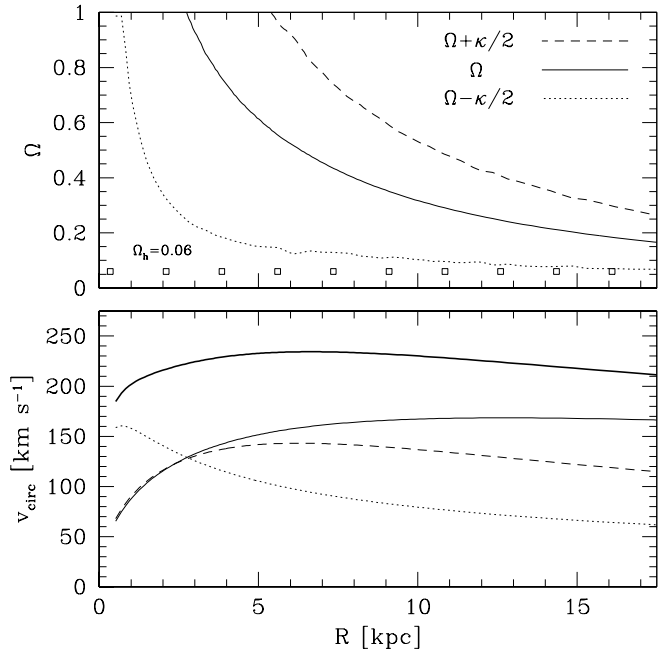}
\caption{\label{curves.1} {\it Top Panel}: The $\Omega$, $\Omega +
\kappa/2$, and $\Omega - \kappa/2$ resonances are plotted against
radius for the fiducial model. The square-dotted line at the bottom of
this plot marks the halo rotation rate. This model has an inner
Lindblad resonance near 15 kpc. {\it Bottom Panel}: The total circular
velocity of the fiducial model is plotted with the thick solid
line. The constituent parts of the galaxy are plotted with the dotted
line (the bulge), the thin solid line (the halo), and the dashed line
(the disk).}
\end{figure}

\section{Simulating a Milky Way Galaxy}

\subsection{Parameters of the Fiducial Simulation}

\begin{figure}[t]
\vspace{6.0cm}
\includegraphics{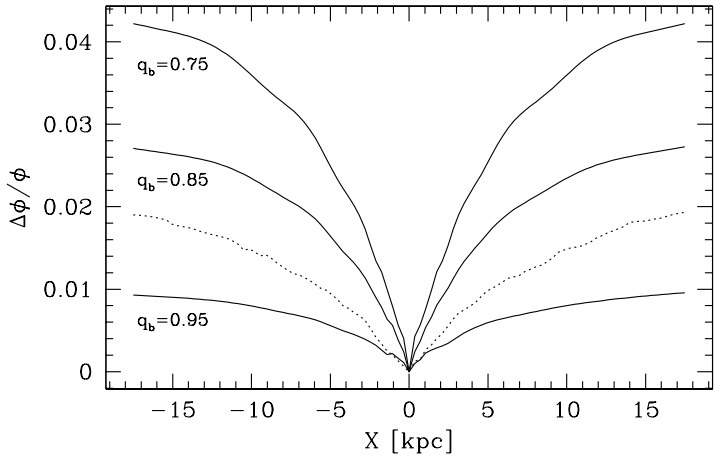}
\caption{\label{potdiff} The fractional difference in the halo
potential between the $x$ and $y$ axes is plotted against the $x$
axis. The three solid lines show halos with different values of
$q_b$. The dotted line plots the initial potential asymmetry of the
entire fiducial model, disk, bulge and halo with $q_b=0.85$.}
\end{figure}

Table 1 lists the physical parameters of the fiducial
simulation. Scaling our $N$-body units to physical values reasonable
for the Milky Way galaxy as in Hernquist (1993), we get $h=3.5$ kpc
and $M_d=5.6\times 10^{10} M_\odot$, making the unit of time 13.1
Myr. With these scalings, the value of the rotational speed at the
solar radius, $R_\odot=2.43h$, is nearly 220 km s$^{-1}$. The fiducial
halo rotation rate, \omh\ =0.06, is equivalent to a pattern speed of
4.5 \kmskpc\ in these unit. (For comparison, the simulations of Bekki
\& Freeman (2002), in which a massless gas disk is embedded in a
rotating triaxial halo, use a pattern speed of 3.8 \kmskpc.) Figure
\ref{curves.1} shows the contributions of each galaxy component to the
total circular velocity. Figure \ref{curves.1} also shows the values
of the $\Omega$, $\Omega+\kappa/2$, and $\Omega-\kappa/2$ resonances as a
function of radius. Although the simulations of Dubinski (1994)
suggest that the equatorial plane of the halo can be as elliptical as
$b/a=q_b=0.7$, we chose a more moderate value of $q_b=0.85$ for our
fiducial simulation. The minor axis ratio, $c/a=q_c$, was set to 0.7.
Figure \ref{potdiff} plots the fractional difference in the potential
along the $x$ and $y$ axes of halos with different values of
$q_b$. The dotted line represents the asymmetry of the potential from
the entire fiducial model.  At five scale lengths, the difference in
the halo potential is $\sim 3\%$, while the difference in the total
system potential is $\sim 2\%$.

\begin{figure}[t]
\vspace{10.0cm}
\includegraphics{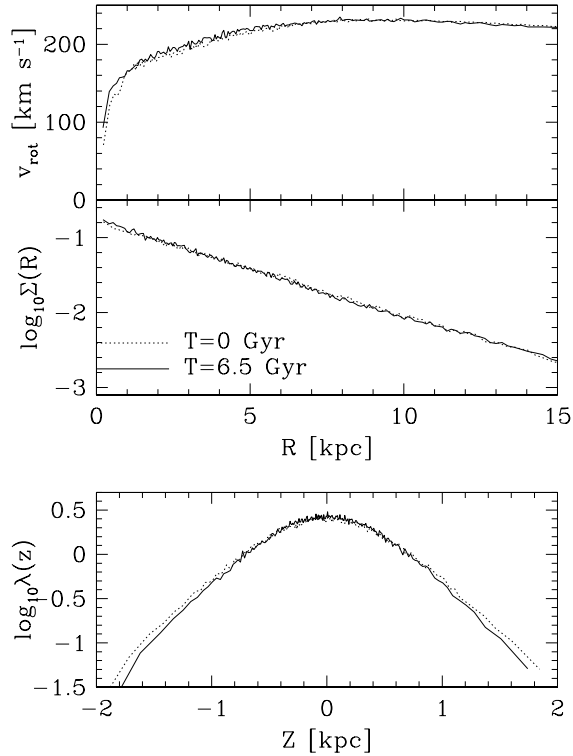}
\caption{\label{sph-stats} Properties of the initial ({\it dotted
line}) and final states ({\it solid line}) of the spherical-halo model
are plotted. The top two panels plot the rotational velocity and the
surface density against cylindrical radius. These statistics were
calculated by binning particles in cylindrical rings of varying width,
but with 500 particles per bin. The bottom panel plots the
vertical linear density $\lambda$ against vertical height $z$. This
was calculated by binning particles by the absolute value of their
$z$-coordinate with 500 particles per bin. }
\end{figure}

To test our numerical method, we ran a simulation with a spherical
halo but still computed the halo's contribution to the potential
through the \scf\ method. Outside of the shape of the halo, which was
set to $q_b=q_c=1$, all the parameters listed in Table 1 were the
same. The simulation was run to a time of 500, which corresponds to a
physical time of 6.55 Gyr. Figure \ref{sph-stats} shows the rotation
curve, the surface density, and the vertical particle density for the
initial and final states of the simulation. These curves show
remarkably little deviation from their initial values, proving that
the integration of the \scf\ into \gadget\ works well for this type of
simulation.

\begin{figure}[t]
\vspace{10.0cm}
\includegraphics{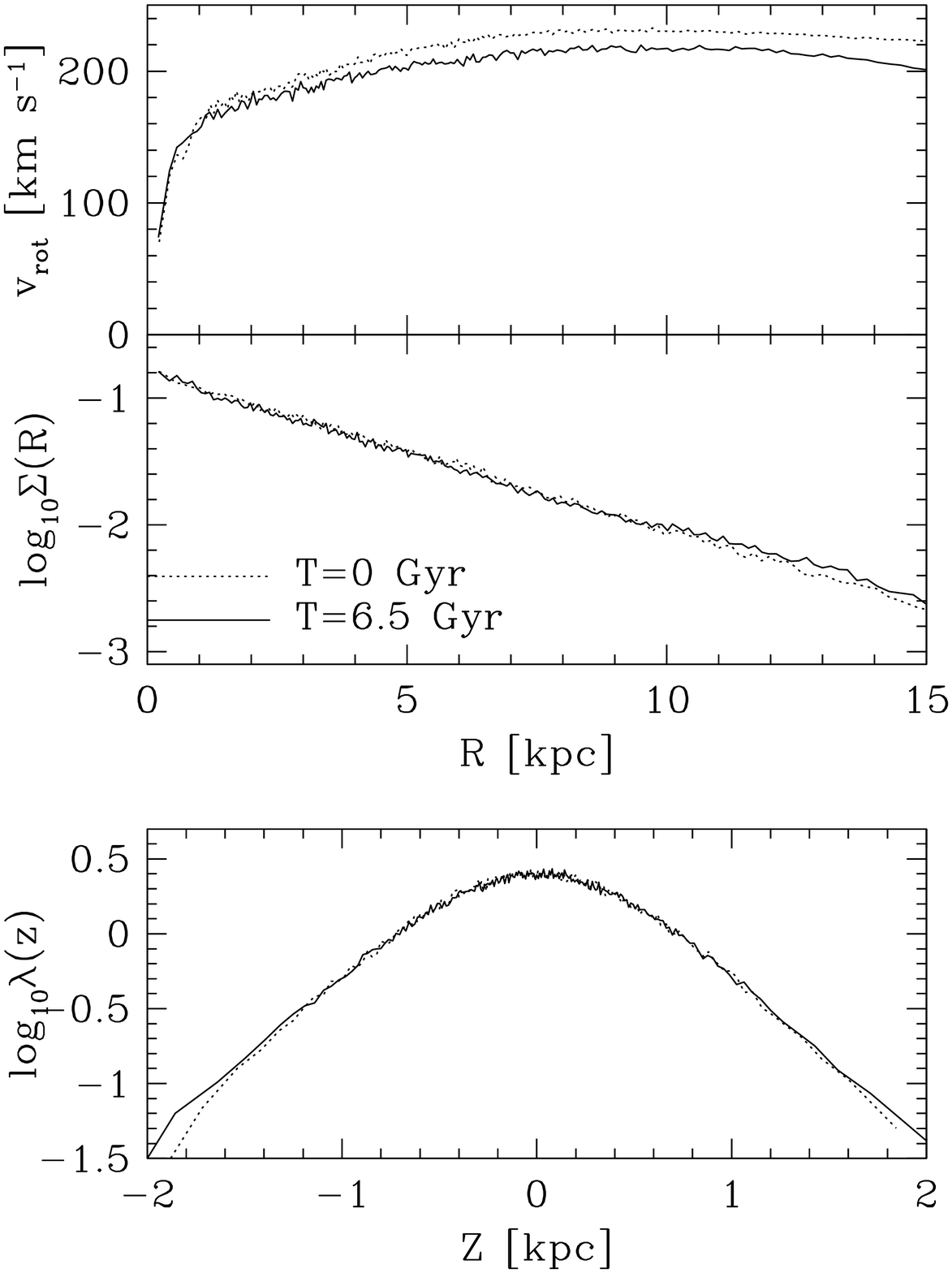}
\caption{\label{fid-stats} Same as Figure \ref{sph-stats}. The halo in
this model is now triaxial, with $q_b=0.85$ and \omh\ = 0.06. There is
little difference between the initial and final states of the
simulation.}
\end{figure}

\subsection{Analyzing the Simulations}

The fiducial model was integrated to a time of 6.55 Gyr as well. The
influence of the triaxial halo of the density structure and rotation
curve can be seen in Figure \ref{fid-stats}. As in the simulation with
the spherically symmetric halo, there is little change in the
azimuthally averaged structure of the disk. The maximum of the
rotation curve is slightly lower than the initial velocity, caused by
a slight heating of the disk as rotational energy is transformed into
thermal. It is interesting to note that the vertical structure of the
disk does not change even after this lengthy integration time. The
disk is not vertically heated by the constant perturbation of the halo
potential.

\begin{figure}[t]
\vspace{9.0cm}
\caption{\label{fid-img} Smoothed particle image of the fiducial
simulation at t=655 Myr. This and other images were created by
smoothing the projected 2-d particle distribution with a Gaussian
smoothing kernel before binning the particles in a 256$\times$256
grid. In order to preserve fine structure while smoothing low-density
regions, the width of the smoothing kernel is different for each
particle, determined by the local particle density.}
\end{figure}

\begin{figure}[t]
\vspace{8.0cm}
\includegraphics{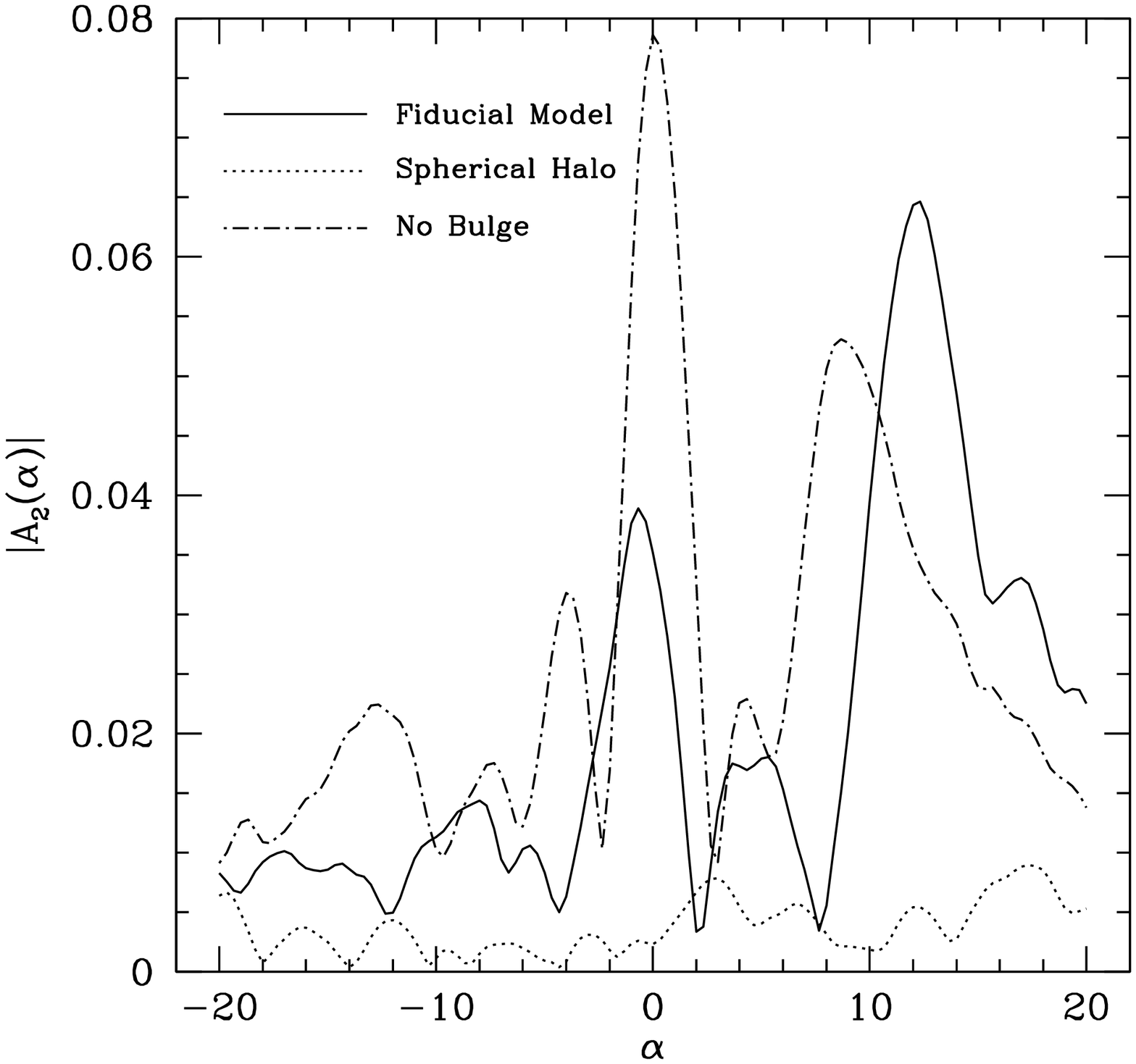}
\caption{\label{fid-four} Fourier decomposition into logarithmic
spirals for several variants of the fiducial model are plotted. The
solid line shows the Fourier spectrum for fiducial model. The
dash-dotted line shows the spectrum for the fiducial model without a
bulge, which shows an enhanced bar component at $\alpha=0$. The
Fourier decomposition of the spherical-halo model is plotted with the
dotted line, which returns just noise. All spectra shown are for
$t=655$ Myr.}
\end{figure}

Structure in the disk is generated by the influence of the halo. A
time-varying, non-axisymmetric potential creates a torque on the disk,
and it is expected that symmetric, $m$=2 structures should
form. Figure \ref{fid-img} shows an image of the fiducial model at a
time of 655 Myr. The bisymmetric structure is apparent. To quantify
the response of the disk to the halo potential, we performed a Fourier
decomposition of the surface density of the disk into logarithmic
spirals (Sellwood \& Athanassoula 1986; Binney \& Tremaine 1987) of
the form $\alpha u+m\phi=ct$, where $u=\ln R$ and $\phi$ is the
azimuthal angle. The amplitude $A$ of each Fourier component is the
Fourier transform of the surface density $S(u,\phi)$;

\begin{equation}
A_m(\alpha)=\frac{1}{(2\pi)^2}\int_{-\infty}^{\infty}du\int_{0}^{2\pi}S(u,\phi)\mbox{e}^{i(\alpha
u+m\phi)}d\phi.
\end{equation}

\noindent In practice, since an $N$-body simulation is resolved by discrete
particles, the integral signs can be replaced by a summation over all
particles and equation (11) then becomes

\begin{equation}
A_m(\alpha)=\frac{1}{N}\sum_{j=1}^{N}\quad{e}^{i(\alpha u_j+m\phi_j)}.
\end{equation}

\noindent The parameter $m$ was set to 2 to look for bisymmetric
structure, and coefficients were calculated for values of $-25\le
\alpha \le 25$. The decomposition of the fiducial model at 655 Myr is
shown in Figure \ref{fid-four}. The decomposition of the model with
the spherical halo at the same time interval is shown for
comparison. Power at negative values of $\alpha$ corresponds to leading
spiral waves. Power at positive values of $\alpha$ corresponds to
trailing spiral waves, which in the fiducial model have a distinct
peak at $\alpha=12$. The pitch angle $p$ for this logarithmic spiral
can be calculated from $p=\tan^{-1}(m/\alpha)=9.5^\circ$. Smaller
values of $\alpha$ correspond to more loosely-bound spiral patterns,
or arms with a larger pitch angle. The peak seen in Figure
\ref{fid-four} centered near $\alpha=0$ is a linear bar structure.

The presence of a bulge in the fiducial model prevents a significant
bar from forming. The peak of the power at $\alpha_{max} = 12$ is
significantly higher than the power at $\alpha = 0$. An identical
simulation to the fiducial model was performed, with the only change
being the removal of the bulge component of the galaxy. The Fourier
decomposition for this bulgeless model at $t=655$ Myr is also plotted
on Figure \ref{fid-four}. The amplitude of the trailing arms is
slightly reduced and \amax\ is at smaller $\alpha$, but the bar
component of the disk has been greatly enhanced. The Fourier
decomposition for the simulation with a spherical halo is also plotted
for comparison. The spectrum of this model shows no significant peaks
and is at low amplitude.

\begin{figure}[b!]
\vspace{7.0cm}
\includegraphics{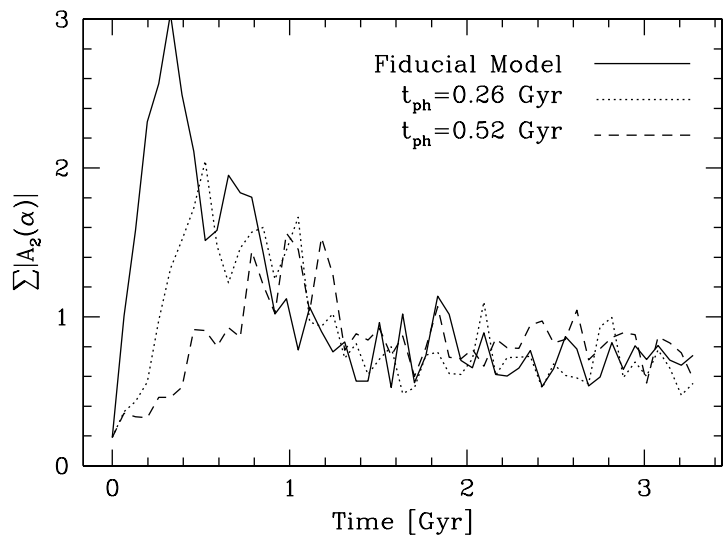}
\caption{\label{fid-arms} The strength of the spiral response of the
fiducial model is plotted against time ({\it solid line}). The {\it dotted}
and {\it dashed} lines represent models where the potential of the
triaxial halo is phased in gradually over time. Phase times $t_{ph}$
are given in the plot. }
\end{figure}

\subsection{Growth of the Spiral Modes}

Because the disk consists solely of collisionless particles, it is
expected that the spiral modes excited by the halo should damp away
simply due to relaxation effects. A galactic disk made entirely of
stars could not sustain spiral patterns indefinitely, although since
the potential of such a system is much smoother than our simulations
the relaxation time would be much longer. We quantify ``arm strength''
as the sum of the amplitudes from $0 \le \alpha \le 25$. A strong bar
would systematically bias this definition of arm strength, since
$\alpha=0$ represents linear structures. However, the simulations for
which this calculation was done did not contain strong bars. The
amplitude of the Fourier spectrum at $\alpha=0$ in Figure
\ref{fid-four} is not the dominant peak.  Above $\alpha=25$, which
corresponds to a pitch angle of $4.6^\circ$, the Fourier spectrum is
usually noise.  Figure \ref{fid-arms} plots arm strength against time
for the fiducial simulation. It shows a rapid growth of the spiral
modes, peaking around 0.5 Gyr, then exponentially decaying to an
asymptotic value of around 0.5. The rms value of the arm strength for
the model with the spherical halo is 0.2.

It is important to note that these simulations all begin with the disk
being out of equilibrium with the triaxial halo. Several simulations
were performed to test the response of the disk to a adiabatic
conversion of the halo potential from spherical symmetry to the full
triaxial form. Initially, the model is begun with the halo potential
being spherical. The triaxial potential is phased in by the function

\begin{equation}
a_i=-(1-\xi)\frac{GM_h}{(r_i+r_0)^2}-\xi\sum_{n,l,m}A_{nlm}\nabla\Phi_{nlm}
\end{equation}

\begin{equation}
\xi=\left(\frac{t}{t_{ph}}\right)^2\left(3-2\frac{t}{t_{ph}}\right)
\end{equation}

\noindent while $t < t_{ph}$. This function has zero derivatives at
$t=0$ and $t=t_{ph}$, so that the change in potential is smooth at the
limits. Simulations were run with $t_{ph}=$ 20 and 40 in $N$-body
units, corresponding to physical times of 0.26 and 0.52 Gyr. These
simulations, plotted in Figure \ref{fid-arms}, reach their peak arm
strength near 2$t_{ph}$ and then proceed to decay along the same curve
as the fiducial model. The peak power is reduced monotonically with
$t_{ph}$, from 2.5 to 1.7 and 1.2. All the curves in Figure
\ref{fid-arms} begin to overlap at a time of 0.8 Gyr and decay to the
same arm strength thereafter. The structure is very similar in all
three simulations at $t \ge 0.8$ Gyr. Even though the peak arm
strength is reduced as the triaxial potential is phased in, the final
state of the disk is unchanged.

\begin{figure}[t!]
\vspace{9.0cm}
\caption{\label{big-img} Same as Figure \ref{fid-img}. The fiducial
model, realized with $10^6$ particles, is shown at $t=655$ Myr. The
scale of the image has been enlarged, with each axis being 15 disk
scale lengths, rather than 10 in Figure \ref{fid-img}. The inner
Lindblad resonance, at 5$h$, is plotted over the image. }
\end{figure}

\begin{figure}[b!]
\vspace{8.0cm}
\includegraphics{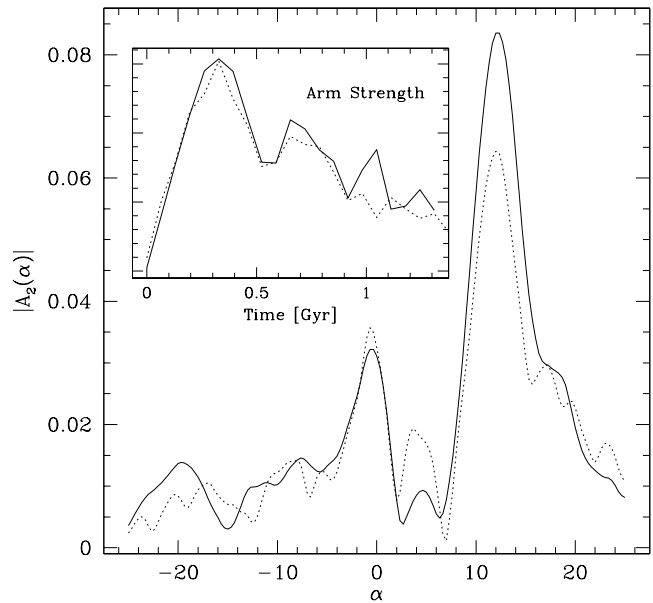}
\caption{\label{big-arms} The Fourier decomposition of the fiducial
model ({\it dotted line}), is compared to that of the simulation with
$N=10^6$ ({\it solid line}) at $t=655$ Myr. The inset box compares the arm strength of
both models. }
\end{figure}

\subsection{Increasing the Simulation Resolution}

To explore the $N$-dependence of our simulations, we increased the
resolution of our fiducial model by an order of magnitude, realizing
the disk with $10^6$ particles and lowering the smoothing length to
$\epsilon = 0.027$. In comparison to the $N=10^5$ simulation, the
structures in the disk, shown in Figure \ref{big-img}, are extremely
apparent and well-resolved. Inside three scale lengths, strong $m=2$
spiral arms are seen. These arms are encircled by two different
structures at different radii. These are caustic-like structures,
where the particle orbits from different radii have crossed each
other, leading to a local density enhancement. The outer caustic is
highly elliptical and is not continuous with azimuth. It has an axis
ratio of 0.7 and a semimajor axis of 6.8$h$. The inner caustic, which
nearly connects with the spiral pattern inside it, has an axis ratio
of 0.9 and a semimajor axis of 3.9$h$. The inner Lindblad resonance
for this model is shown on top of the image in Figure
\ref{big-img}. Located just outside the inner caustic, at 5$h$, it
connects the inner and outer caustics at the minor axis of the outer
caustic.

\begin{figure}[t!]
\vspace{7.8cm}
\includegraphics{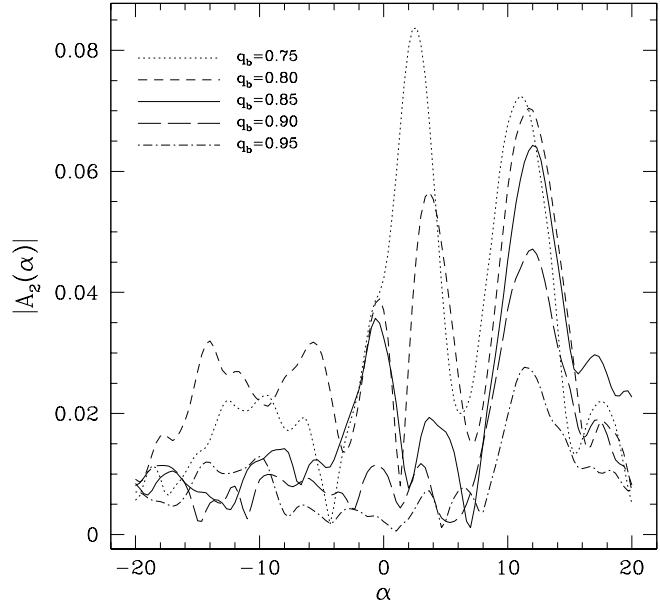}
\caption{\label{q-results} The Fourier decompositions for five
simulations with different flattenings at $t=655$ Myr. The peak in the spectrum at
$\alpha=12$ increases as the halo becomes more flattened. At
$q_b<0.85$, features at smaller $\alpha$ become comparable to the peak
at $\alpha=12$. }
\end{figure}

\begin{figure}[b!]
\vspace{7.1cm}
\includegraphics{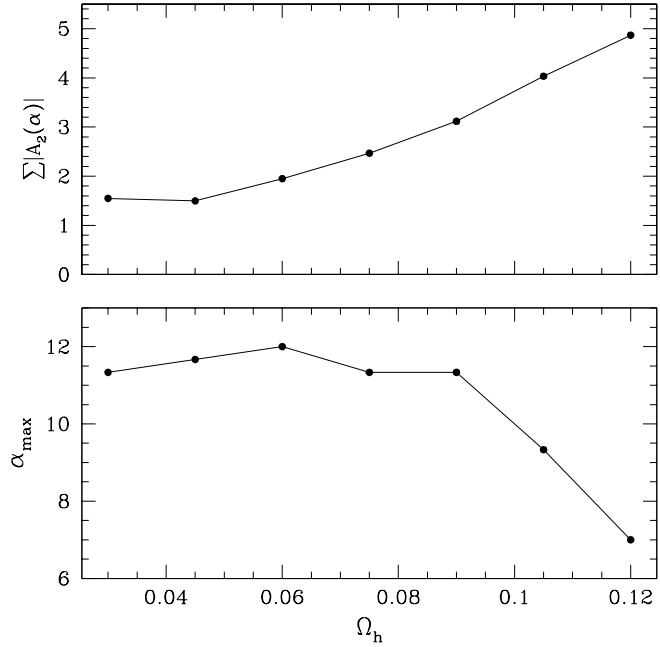}
\caption{\label{om-results} {\it Top Panel}: The arm strength of the
disk, defined as the sum of the Fourier coefficients between $0 \le \alpha
\le 25$, is plotted against the halo rotation rate, \omh. {\it Bottom
Panel}: The location of the peak of the Fourier spectrum, \amax, is
plotted against \omh. As the response of the disk increases, the
peak in the spectrum moves to lower $\alpha$, logarithmic spirals with
higher pitch angle. All calculations were done at $t=655$ Myr.}
\end{figure}

Figure \ref{big-arms} shows the Fourier decomposition of the $N=10^6$
simulation in comparison to the fiducial model. The peaks in the power
of both simulations are at the same locations, with a slightly higher
peak at $\alpha=12$ and lower peak at $\alpha=0$ for the higher
resolution simulation. Figure \ref{big-arms} also plots the arm
strength as a function of time. The peak arm strength of the high
resolution simulations is slightly higher than that of the fiducial
simulation, and the two curves decay at nearly the same rate. These
results demonstrate that the fiducial resolution is adequate for the
analysis performed.

\subsection{Altering Halo Parameters}

Cold dark matter simulations do not predict a single shape to dark
matter halos; rather a distribution of axis ratios and
triaxialities. For the halo figure rotation, the fiducial value of
$\Omega_h=0.06$ is somewhat arbitrarily chosen and limits on this
parameter are not defined. We ran simulations to test the effects of
varying these parameters, first varying $q_b$ with \omh\ held
constant, then varying \omh\ with $q_b$ held at 0.85.

Figure \ref{q-results} shows the Fourier decompositions of five
models, including the fiducial model, with different levels of
flattening. The most extreme model has $q_c=0.5$ and $q_b=0.75$. The
most moderate model has axis ratios of 0.9 and 0.95. (The value of
$q_c$ for our models is always $2q_b-1$.) All the models show a peak
near $\alpha=12$, showing that the pitch angle of the spiral arms
created remains nearly the same in these different models. As $q_b$
gets closer to unity, the response of the disk falls off; at
$q_b=0.95$, the amplitude of the Fourier peak is a factor of 2.3 lower
than the fiducial model, and the disk itself shows little visible
evidence for spiral structure. As $q_b$ gets smaller, however,
response of the disk around $\alpha=0$ increases dramatically. In the
most flattened model, the amplitude at $\alpha \sim 3$ is greater than
that at $\alpha=12$.

\begin{figure}
\epsscale{1.0}
\caption{\label{4disk-om} Images are shown for four values of \omh:
0.045, 0.075, 0.09, and 0.12. The increased response of the disk, as
well as the tighter spiral pattern, can be seen as the halo rotation
rate is increased. The images represent the simulations at a time of
655 Myr. }
\end{figure}

Figure \ref{om-results} shows how the integrated arm strength and the
value of \amax\ each change with the rotation rate of the halo,
\omh. As the halo rotation increases, the response of the disk is
monotonically increasing when \omh\ $\ge 0.045$. As \omh\ increases to
\omh\ = 0.12, the value of \amax\ decreases from 12 to 7. This can be
seen in Figure \ref{4disk-om}, which shows images of four of these
models.

\section{Exploring Parameter Space}

The mass distribution of the Milky Way, although studied for several
decades, is not known with certainty. The mass distribution of other
galaxies are not known with much confidence at all, although it is
known that there exists a range of halo and disk masses. Low surface
brightness galaxies are thought to be dark-matter dominated, while
many researchers espouse the maximal disk hypothesis presented by van
Albada \& Sancisi (1986) for galaxies with high surface brightness

\begin{figure}[t]
\vspace{6.0cm}
\includegraphics{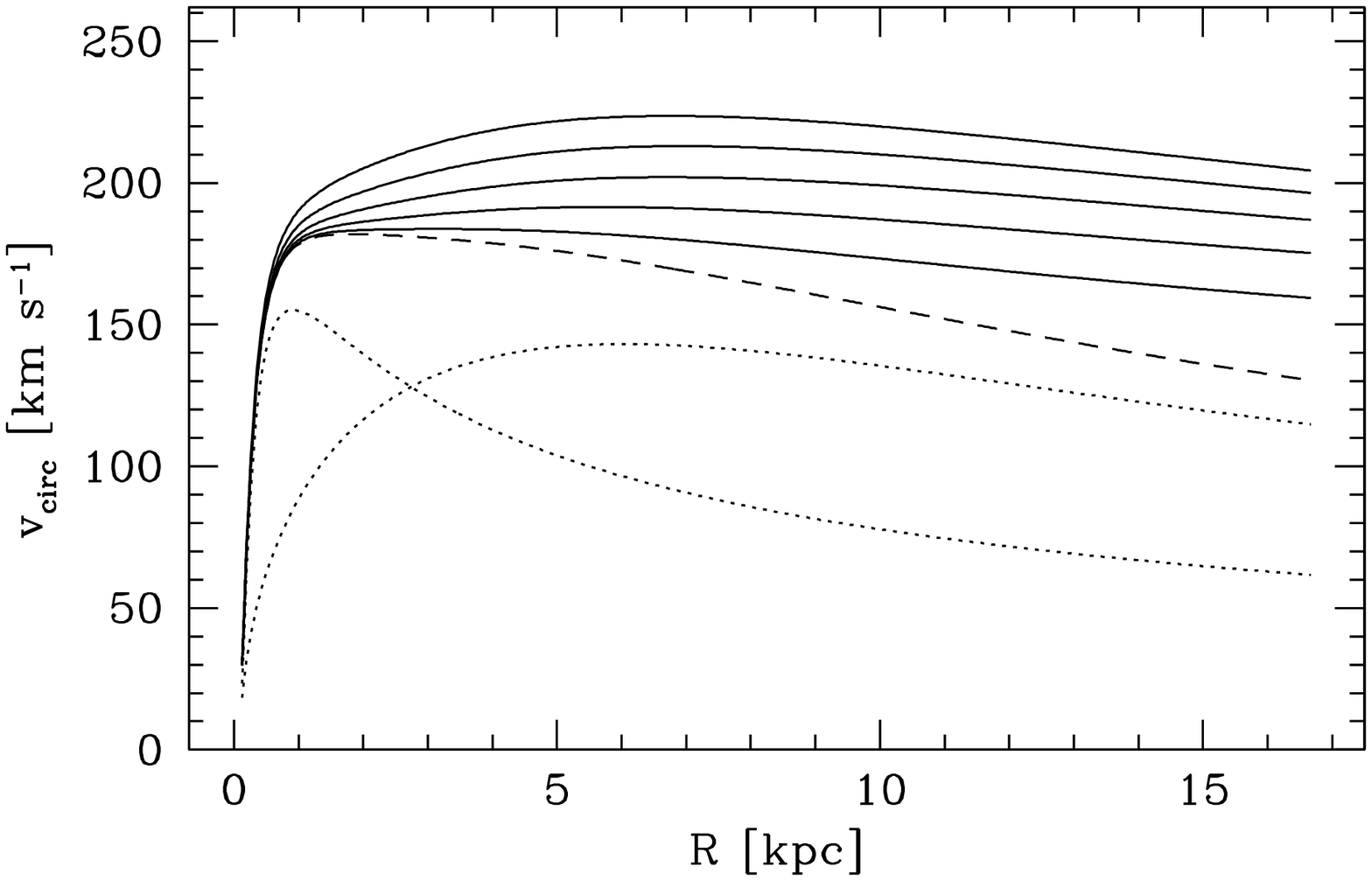}
\caption{\label{rot-curvesD} Circular velocity curves for models
D1--D5 (disk-dominated) are plotted. The dotted lines represent the
contributions to the circular velocity from the disk and the
bulge. The dashed line is the combined circular velocity of the disk
and bulge. The solid lines are the total circular velocities of the
five models. The lowest curve, which is closest to a maximal-disk
model, is D5. The highest curve is model D1. The specific halo
parameters used to create these models are listed in Table 2. }
\end{figure}

\begin{figure}[t]
\vspace{6.0cm}
\includegraphics{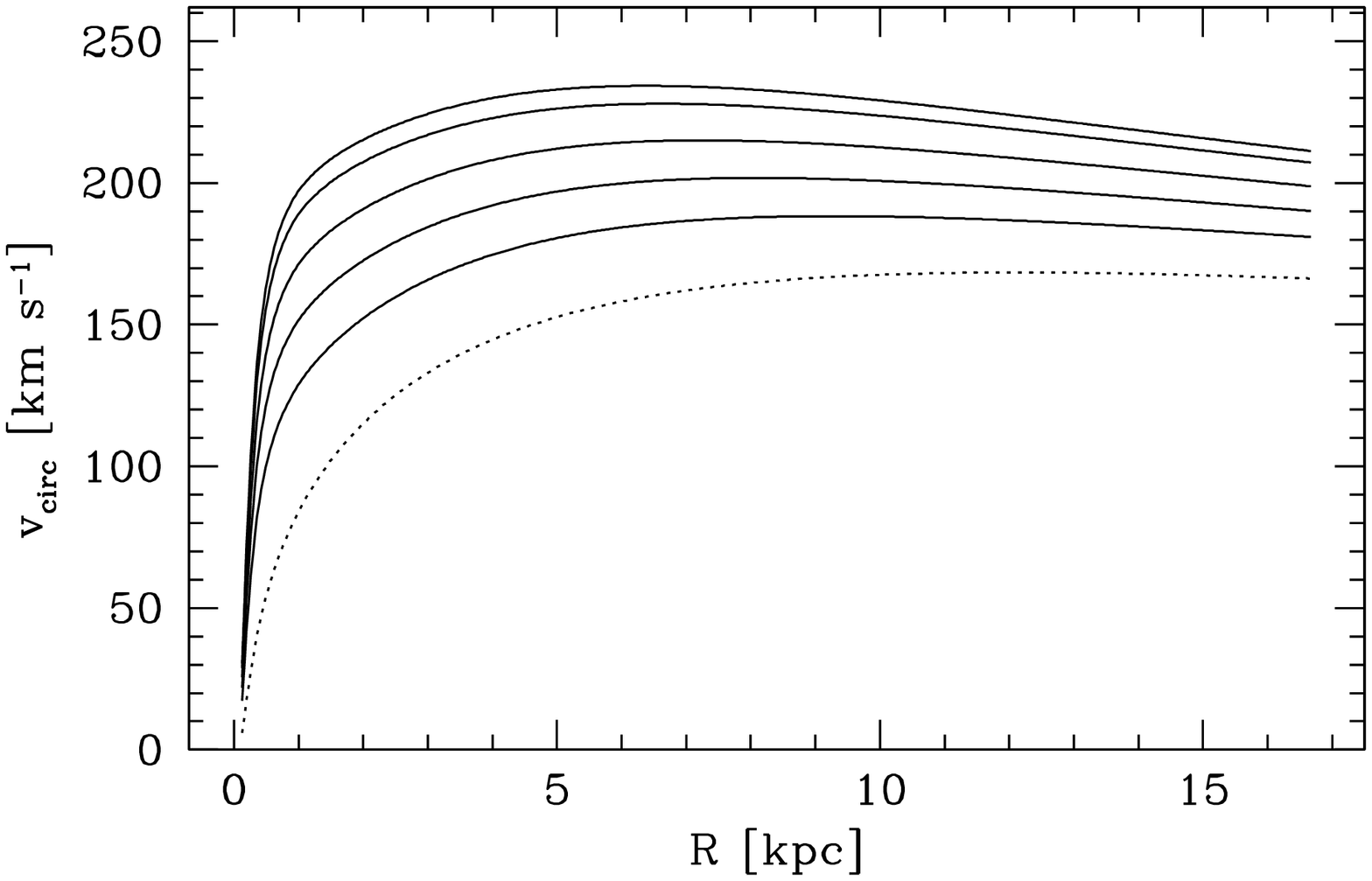}
\caption{\label{rot-curvesH} Circular velocity curves for models
H1--H5 (halo-dominated) are plotted. The dotted line represents the
circular velocity of the halo, which is constant in all these
models. The solid lines are the total circular velocity from the disk,
halo and bulge. Model H1, with a disk mass of 0.3, is the lowest
curve; Model H5, which is the fiducial model, is the highest. For each
model, the bulge mass $M_b=0.3M_d$. }
\end{figure}

To explore these parameters, we constructed several sets of models in
which all three components of a disk galaxy--disk, bulge, and
halo--were varied in systematic ways.  First we explored models where
the disk mass was reduced and the dynamics became more heavily
influenced by the potential field of the halo. Second, we ran models
where the parameters of the halo, namely the interior slope $\gamma$
and the scale radius $r_0$, were altered such that the mass
distribution of the galaxy became disk-dominated at small radii. The
total mass of the halo is constant, or nearly so, if one considers the
truncation radius of the halo to be at a value of 100$h$, the value
chosen to calculate the moment of inertia of the halo. We characterize
each of these models by its {\it mass ratio}, $(M_d/M_h)_{R_\odot}$,
the ratio of the disk to halo mass within the solar radius, 8.5
kpc. Table 2 lists the general properties of the disk-dominated
models, denoted D1--D5, where D5 is the most nearly maximal disk with
a mass ratio of 4.53, and the properties of the halo-dominated models,
denoted H1--H5, where H1 has the least massive disk with a mass ratio
of 0.22. Figures \ref{rot-curvesD} and \ref{rot-curvesH} show the
circular velocity profiles for the halo- and disk-dominated models
respectively. In all models the bulge mass is set to 0.3$M_d$.

\begin{deluxetable}{ccccccc}
\tablecolumns{7} 
\tablewidth{32pc} 
\tablecaption{Properties of the Models with Different Mass Distributions} 
\tablehead{ \\
\multicolumn{3}{c}{Halo-Dominated Models} &
\multicolumn{4}{c}{Disk-Dominated Models} \\ \\ \hline \\
\colhead{Model Name} & \colhead{($M_d$/$M_h$)$_{R_\odot}$} &
\colhead{$M_d$} &
\colhead{Model Name} & \colhead{($M_d$/$M_h$)$_{R_\odot}$} &
\colhead{$\gamma$} & \colhead{$r_0$}
}
\startdata

H1 & 0.22 & 0.3 & D1 & 0.91 & 0.8 & 3.63 \\
H2 & 0.38 & 0.5 & D2 & 1.13 & 0.6 & 3.73 \\
H3 & 0.53 & 0.7 & D3 & 1.51 & 0.4 & 3.98 \\
H4 & 0.68 & 0.9 & D4 & 2.27 & 0.2 & 4.48 \\
 H5$^\ast$ & 0.76 & 1.0 & D5 & 4.53 & 0.0 & 5.69 \\

\enddata
\tablecomments{\footnotesize ($^\ast$) Model H5 is the fiducial model described
in \S 3}
\end{deluxetable}

Third, we created a set of models in which the disk and halo mass
distributions were held constant and the bulge mass and scale length
were varied. The disk mass was $M_d$ = 0.8 and the halo parameters
were the same as the fiducial model. The bulge masses were $M_b$ = 0,
0.1, 0.3, 0.5, 0.7, 0.9, 1.3, and 2.0. The value of each $r_b$ was
calculated such that the average density inside one bulge scale radius
was the same as that of the fiducial bulge, namely

\begin{equation}
r_b=r_{b,0}\left(\frac{M_b}{M_{b,0}}\right)^{1/3}
\end{equation}

\noindent where $r_{b,0}$=0.2 and $M_{b,0}=0.3$.

Our last set of models centered on models that had lower rotation
velocities, values closer to 100 \kms\ than to the fiducial value
of 220 \kms. The motivation behind these models will be discussed in
\S6. The specific parameters used are as follows: The halo parameters
are identical to the halo in model D5, which is the model with the
lowest central concentration. The disk mass was 0.5. The bulge mass
was varied, having values of 0, 0.075 and 0.15.

\begin{figure}
\epsscale{1.0}
\caption{\label{4bar} Images of four simulations at increasing levels
of bar strength, \blah2max. The values of \blah2max\ are printed above each
panel. In contrast to the fiducial resolution, all simulations
used in this figure has a resolution of $3\times 10^5$. The higher
resolution was used to reduce noise which can be amplified in longer
simulations. The statistics calculated in Figures \ref{bar.b3} and
\ref{bar.b15} used higher resolution simulations.}
\end{figure}

The results of these simulations will be discussed in the following
two sections, focusing on the formation of bars and morphology of the
spiral arms induced in the disk by the halo.

\section{Bar Formation}

Bars are simple to detect in our simulations, since they represent the
$\alpha=0$ modes of the Fourier decompositions. Models with strong
spiral arms, however, can also return high amplitudes at $\alpha=0$
compared to models with no structure at all. To prevent false
detection of bars, we followed the procedure described in Efstathiou
\etal\ (1982) to quantify the strength of the bar. The particles are binned
in radial rings of linear separation and the amplitude and phase of
the $\alpha=0$, $m=2$ component is calculated for each ring. If a bar
is present, it is defined as existing within the rings that have
phases which are coherent to within 10\%. The bar strength is the
maximum Fourier component within this coherence and is denoted by the
symbol $\delta_2^{max}$. The rings are 0.1$h$ in width and contain
several thousand particles each. To be considered a valid detection,
the coherence must be at least 5 rings in length but does not have to
begin at the center-most ring-- some simulations with obvious bars do
not begin having coherent phases until $R\sim 0.3-0.5h$. From this
method we can determine the amplitude, length, and pattern speed of
the bar. The binning of particles in cylindrical rings eliminates
false detection of bars in simulations with strong spiral patterns
because these models will have high values of \blah2max\ but will not
have coherent phases.

\begin{figure}
\epsscale{0.8}
\plotone{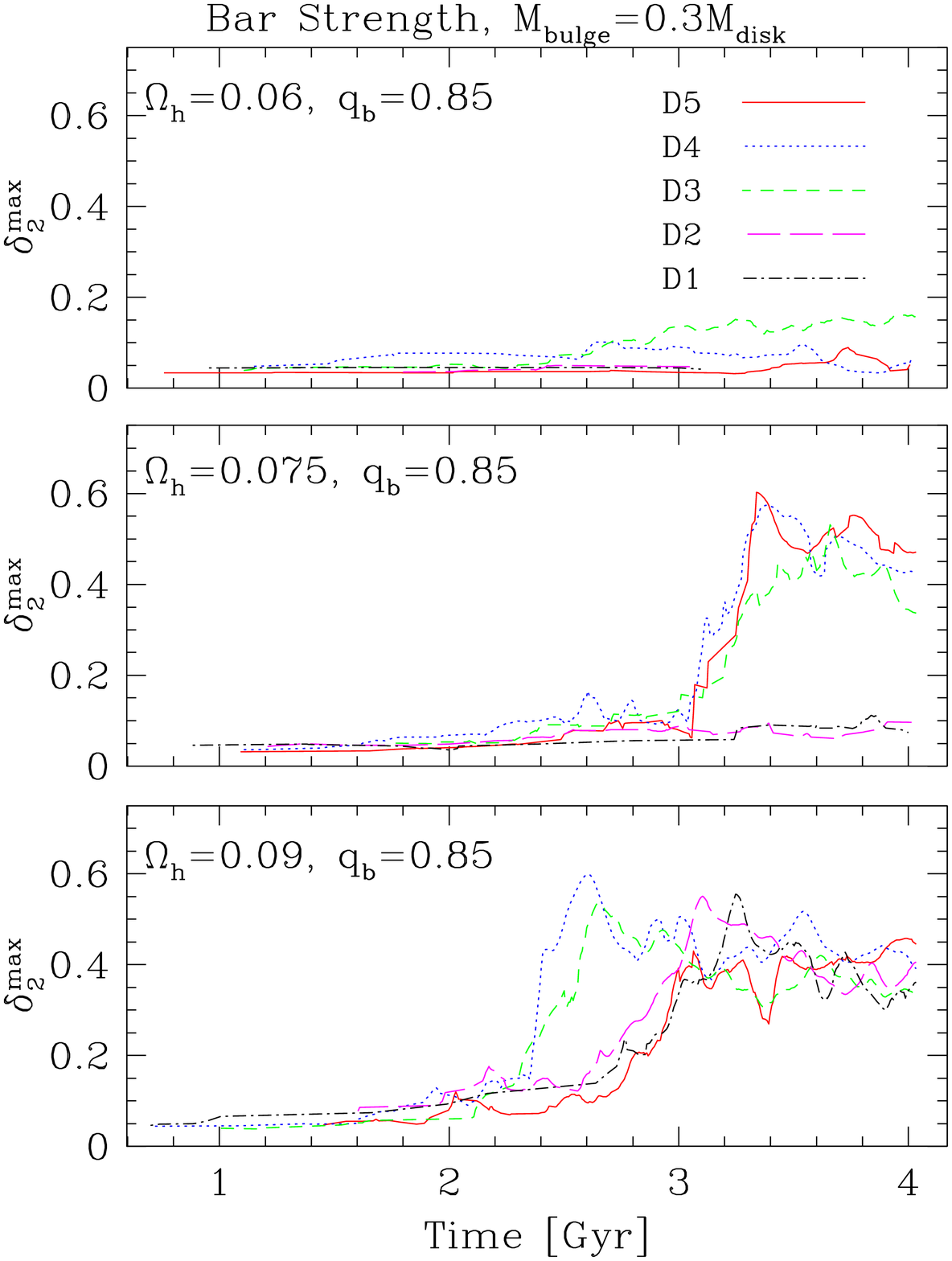}
\caption{\label{bar.b3} The bar strength for models D1--D5 are plotted
as a function of time at three different values of \omh. }
\end{figure}

\begin{figure}
\epsscale{0.8}
\plotone{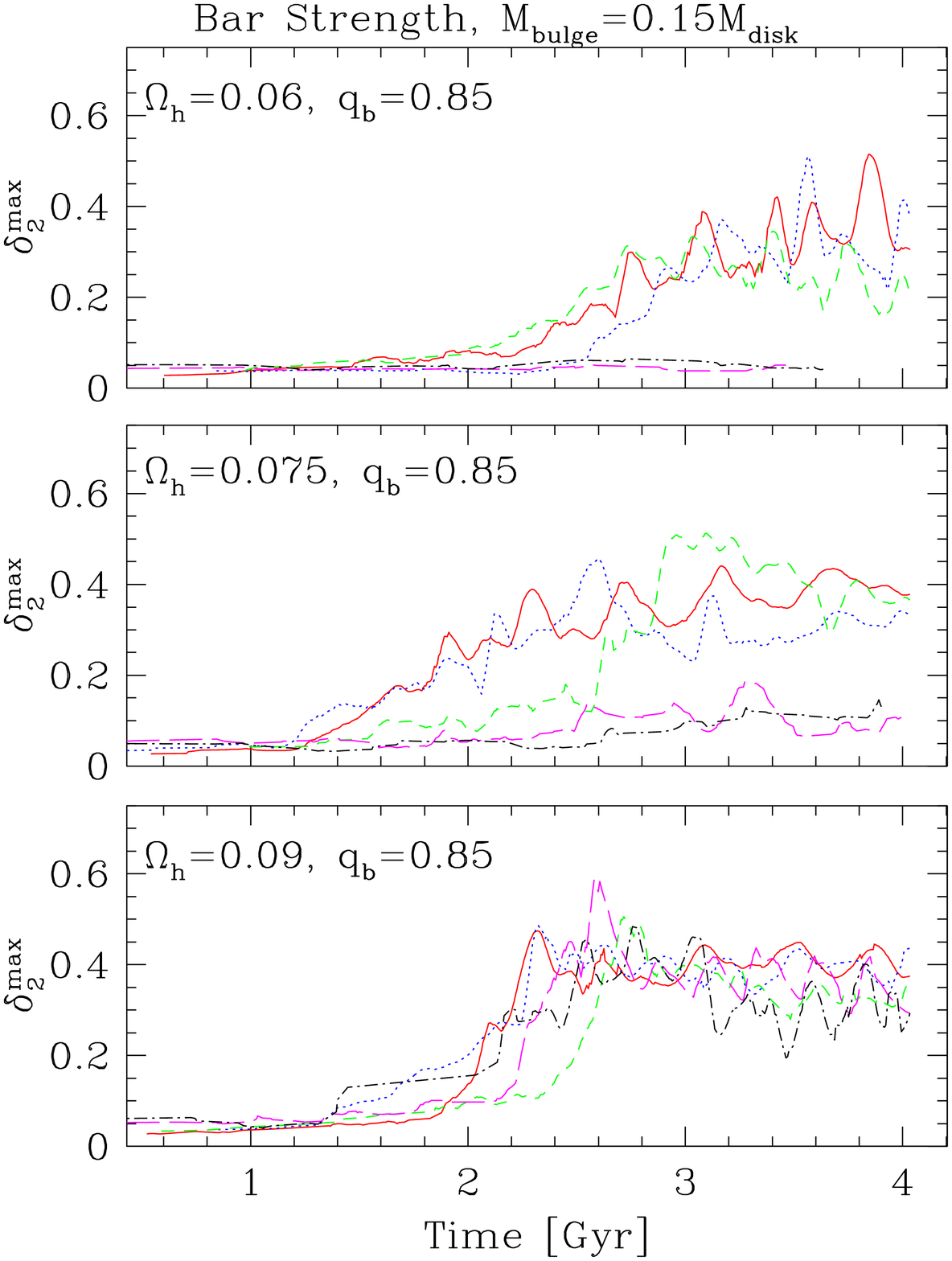}
\caption{\label{bar.b15} Same as Figure \ref{bar.b3}. The bulge mass
for these models has been reduced to 0.15 to make the models less
stable against bar formation. The different line styles represent the
same models as in Figure \ref{bar.b3}. }
\end{figure}

Figure \ref{4bar} shows simulations at four different levels of bar
strength: \blah2max\ = 0.1, 0.2, 0.4, and 0.6. Values of \blah2max\ less
than or equal to 0.2 have weak bars that are generally $\sim 1$ disk
scale length in size. At \blah2max\ = 0.4, the bar is very distinct and
longer. At \blah2max\ = 0.6, the highest value seen in our simulations,
the bar dominates the mass of the disk and can be nearly four scale
lengths in size. When integrated with spherical halos, all of the
models presented in this section developed no detectable bars.

Figure \ref{bar.b3} shows the time evolution of bar strength in models
with different values of halo rotation rate. For the fiducial value of
\omh$=0.06$, the disk-dominated models are generally stable. Only
model D3 develops a detectable bar, one which asymptotes to
\blah2max$\sim 0.15$. When \omh\ is increased, however, bar instabilities
become much larger. At \omh$=0.075$, the models with the lowest mass
ratios (D1 and D2) are stable, but the the other models show
significant bar formation. The growth in the bar modes is very gradual
at first, but at 3 Gyr models D3, D4, and D5 rapidly grow their bars,
which peak at values of \blah2max\ between 0.5 and 0.6 and slowly decline
in amplitude.  In the bottom panel of Figure \ref{bar.b3}, \omh\ has
been increased to 0.09. All the models begin to show bar structures at
$\sim$ 2.5 Gyr and become bar-dominated by 3 Gyr. Models D3 and D4
make a rapid transformation to being bar-dominated, while \blah2max\ for
the other models grow more gradually. Regardless of the time it takes
for the bar to grow, the values of \blah2max\ for all the models is close
to 0.4 once the bar has reached its peak strength. The values of
\d2ave, the bar strength averaged over the last 0.8 Gyr of the
simulation, do not show any correlation with mass ratio. For model D1,
\d2ave\ = 0.39, while for model D5 \d2ave\ = 0.40. The halo-dominated
models, including the fiducial model H5, are all stable against bar
formation, even with a halo rotation rate of 0.09.

As stated above, the stability of the disks shown in Figure
\ref{bar.b3} is in part due to the mass of the bulge. We decreased the
value of the bulge mass to $0.15M_d$ and re-ran our disk-dominated
models to investigate bar formation in galaxies with less prominent
bulges. Figure \ref{bar.b15} shows the results. At \omh\ = 0.06,
models D3--D5 develop distinct bars while the other models remain
stable. At \omh\ = 0.075, all the models show some level of bar
strength, with the average value of \blah2max\ loosely correlating with
increasing mass ratio. At \omh\ = 0.09, all the models show rapid bar
growth and become bar-dominated between 2 and 2.5 Gyr. The values of
$\langle \delta_2^{max}\rangle$ also correlate roughly with mass
ratio. For model D1 $\langle \delta_2^{max}\rangle=0.30$, increasing
to $\langle \delta_2^{max}\rangle=0.41$ for model D5. In these
simulations, the growth of the bar begins earlier but is slower than
the models with $M_b=0.3M_d$. The values of \d2ave\ are slightly higher
in the $M_b=0.3M_d$ simulations for those models which develop bars.

The bar pattern speeds do not appear to correlate with either \omh\ or
\blah2max. The simulations that develop strong bars all have bar
pattern speeds near 0.5 radians per unit time (in the same units as
\omh). The pattern speeds do not change on average once the bar has
reached its peak amplitude. Since our simulations employ a static halo
potential our results do not reflect the dynamical friction and
angular momentum transfer that would surely occur in physical
galaxies, reducing the pattern speed (Debattista \& Sellwood 2000;
Valenzuela \& Klypin 2002).

\begin{figure}[t]
\vspace{10.5cm}
\includegraphics{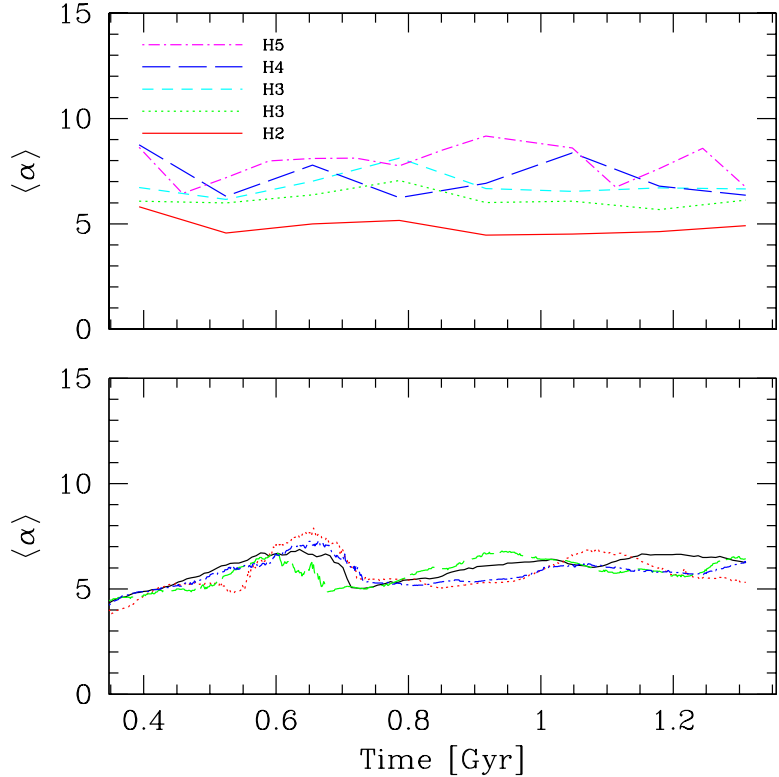}
\caption{\label{pave-t} {\it Top Panel}: The weighted average of the
Fourier spectrum, \aave, is plotted against time for models H1--H5 and
a halo rotation rate of 0.09. The values of \aave\ for each model
remain nearly constant in time for 1 Gyr. {\it Bottom Panel}: \aave\
is plotted against time for four realizations of the low-velocity
model described in \S 4 with $M_b=0.3M_d$. The black line represents a
simulation resolved with $N=3\times 10^5$. The other lines represent
simulations with the fiducial resolution of $10^5$. The time range
shown is the range used when calculating the time average, \aavet,
which is then used to calculate the average pitch angle, \pave. }
\end{figure}

\section{Spiral Morphology}

\subsection{Spiral Arm Pitch Angle}

Although in some simulations the peak in the Fourier spectrum is well
defined and narrow, in others a simple search for the maximum value of
$|A_2|$ can vary significantly in time if there is more than one peak
in the spectrum. Pitch angles in spiral galaxies are known to vary
with radius and are measured as an average value (Kennicutt
1981). Therefore, to quantify the morphology of the spiral structure
in our simulations we calculate an average value of $\alpha$, summed
from $0\le\alpha\le25$ and weighted by the Fourier amplitude at each
value of $\alpha$. The simulations for which this quantity, \aave, were
calculated did not contain bars as defined in \S 5, and so using
$\alpha=0$ as the starting point of the weighted average does not bias
the results to smaller values of \aave.  Only amplitudes greater than
half the rms amplitude in that range are actually counted in the
summation. The value of \aave\ varies in time about a mean value at a
level of $\sim$ 5-20\%. Fourier spectra with higher overall amplitude,
usually found in models with higher values of \omh, have smaller rms
fluctuations. The weighted average of $\alpha$ is again averaged over
the time range 0.4--1.3 Gyr. The lower of these limits is chosen to
allow the disk to reach equilibrium with the halo. As the simulations
progress the arm strength decays, and so the integrations were
stopped at 1.3 Gyr. In our notation, \aave\ is the weighted average of
the Fourier spectrum for a given moment in time, and \aavet\ is the
average of this value over time range mentioned. The quantity \aavet\
can then be converted into an average pitch angle, \pave.

\begin{figure}[t]
\vspace{8.5cm}
\includegraphics{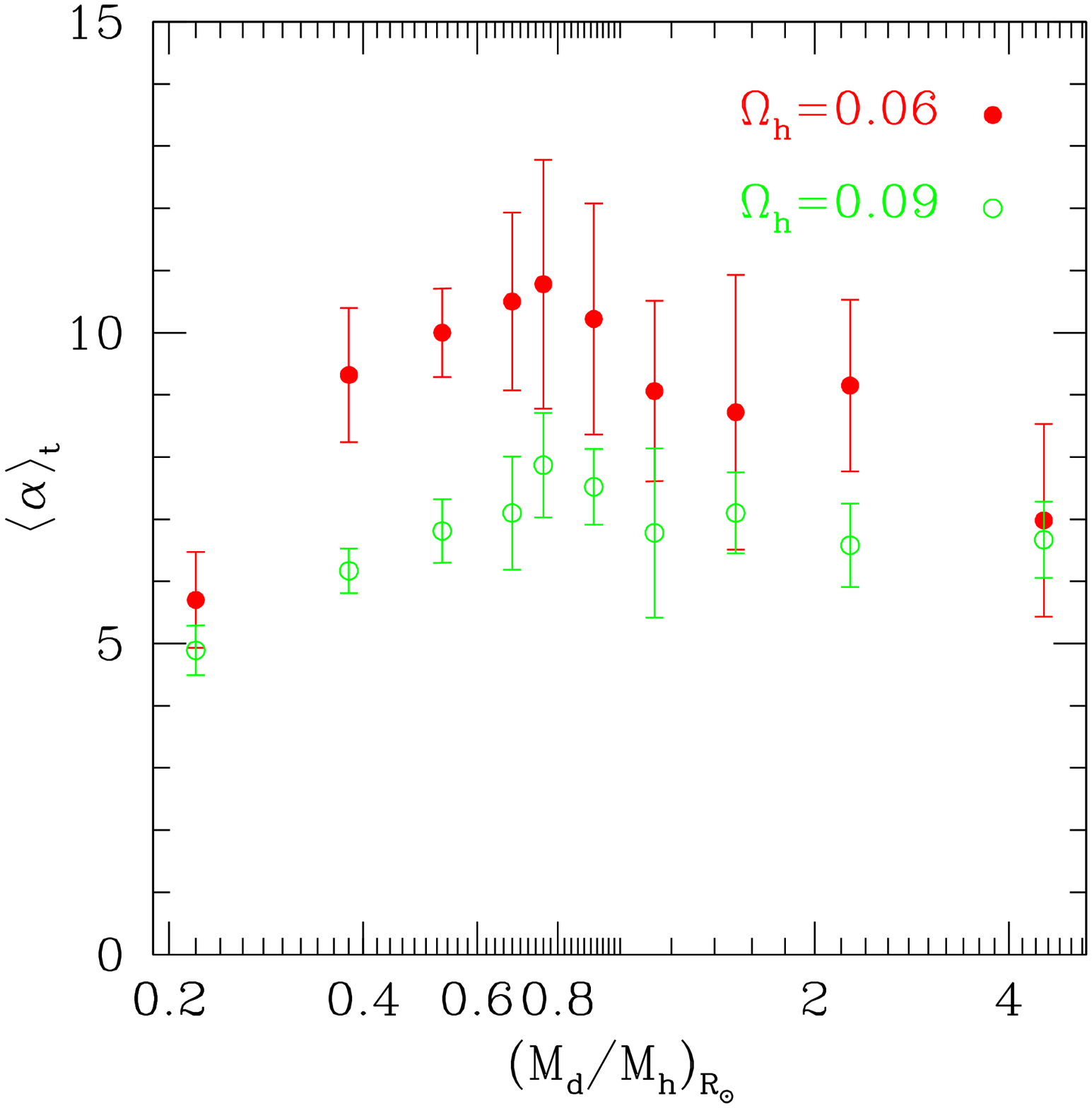}
\caption{\label{aave} The time average of the Fourier spectrum,
\aavet, is plotted against mass ratio for all models listed in Table
2. The different plot symbols represent different halo rotation rates,
0.06 and 0.09. The error bars are standard deviations of the values of
\aave\ averaged to calculate \aavet. The error bars for the simulations
with \omh\ = 0.06 are larger because of the lower overall amplitude of
the Fourier spectra at lower \omh, allowing more noise into the
calculations. }
\end{figure}

\begin{figure}
\epsscale{1.0}
\caption{\label{4disk.09} Images of models H1--H4 at a time of 655
Myr. These simulations all have \omh\ = 0.09. The mass of each disk is
printed above each panel. The mass of the bulge is $M_b=0.3M_d$ for
each model. }
\end{figure}

The top panel in Figure \ref{pave-t} shows the evolution of \pave\
with time for the halo-dominated models H1--H5 with \omh\ =
0.09. Although there is some variation, the average value with time
for each simulation is well-defined. The bottom panel in Figure
\ref{pave-t} shows four realizations of the low-velocity model with
$M_b=0.3M_d$, the last of which was resolved at $3\times 10^5$
particles rather than the fiducial value of $10^5$. The simulations
were sampled with much higher frequency to investigate how \aave\
varies on small time scales. The results show little difference between
all four realizations, showing that one realization at the fiducial
resolution, although not ideal, is adequate for our results.

The clear trend in the top panel of Figure\ref{pave-t} is that \aave\
decreases with decreasing mass ratio. Figure \ref{aave} shows the
value of \aavet\ plotted against mass ratio for all the models listed
in Table 2. The results for \omh\ = 0.09 have smaller error bars than
those for \omh\ = 0.06 because of the higher amplitude of the Fourier
spectra for \omh\ = 0.09. For both sets of simulations, the value of
\aavet\ is maximum at the fiducial mass ratio of 0.76 and decreases as
the mass ratio gets bigger or smaller. As expected from the results of
Figure \ref{om-results}, the data for \omh\ = 0.09 are shifted to
lower values of \aavet\ as compared to the data for \omh\ =
0.06. Figure \ref{4disk.09} shows images of models H1--H4 at 655 Myr
with \omh\ = 0.09. The increased response of the disk with decreasing
mass ratio is clearly evident. The spiral structure for model H2 can
be traced for well over 2$\pi$ radians. The image of model H1, with
the lowest mass disk, resembles a classic Kalnajs disk (\eg\ Binney \&
Tremaine 1987). The caustic structures seen in the high-resolution run
of the fiducial simulation are seen throughout the disk and are highly
elliptical.  Differential rotation has caused the oval caustics to
overlap and create a spiral pattern. 

\subsection{Bulge to Disk Ratio and the Hubble Sequence}

The third set of models described in \S 4, those with varying bulge
mass, can be thought of as an ersatz Hubble sequence, at least along
one dimension of the required criteria. The properties of the disk and
halo were chosen as to produce significant amplitudes in the Fourier
decomposition, reducing noise in the measured value of the pitch
angle.

\begin{figure}
\epsscale{1.0}
\caption{\label{4bulge.105} Images of models with bulge masses
$M_b=$0.1, 0.3, 0.7, and 2.0. For all simulations, $M_d=0.8$ and \omh\
= 0.105. }
\end{figure}

\begin{figure}[t]
\vspace{10.5cm}
\includegraphics{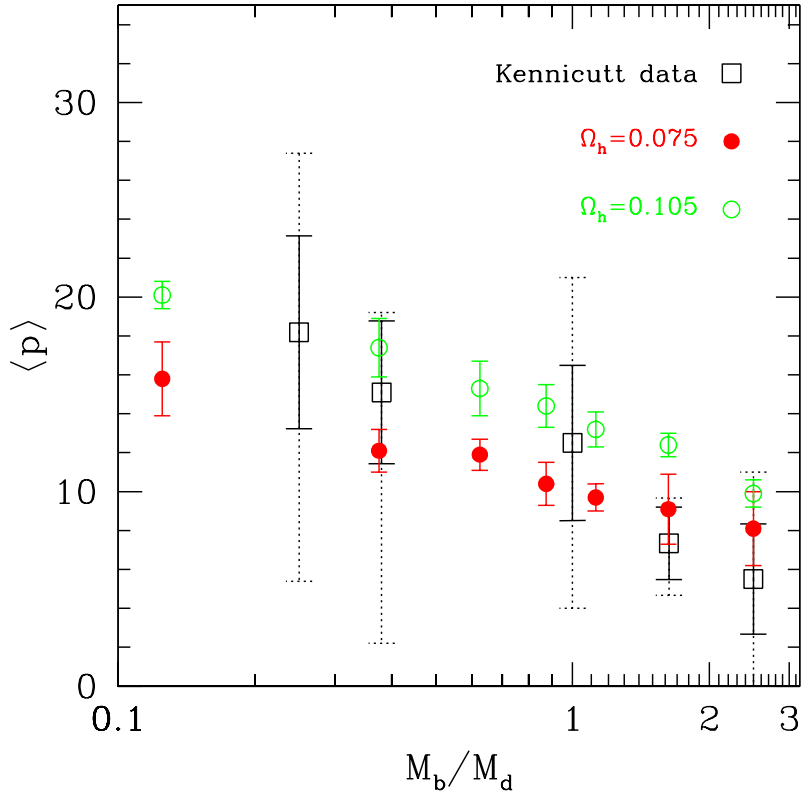}
\caption{\label{hubble} The average pitch angle, \pave, is plotted
against bulge-to-disk ratio for the models described in \S 6.2. The
different solid plot symbols represent different halo rotation rates,
0.075 and 0.105. Error bars are statistical. The open squares represent
data taken from Kennicutt (1981) of pitch angles for Hubble types Sa,
Sab, Sb, Sbc, and Sc. The solid error bars are the standard deviations
of these data. The dotted error bars are the high and low values of
pitch angle for each Hubble type. }
\end{figure}

Figure \ref{4bulge.105} shows images of four models, $M_b=0.1, 0.3,
0.9$ and 2.0 with \omh\ = 0.105. As the bulge mass varies from small
to large, the spiral arms vary from loosely to tightly wound. The arms
themselves vary in thickness; the larger the bulge mass, the narrower
the spiral arms. For the model with $M_b=0.1$, the spiral arms can be
traced to within one scale radius of the center. For $M_b=2.0$, the
arms do not begin until outside two scale radii (although a
high-resolution simulation could reveal structure inside this radius,
as it did for the fiducial simulation). Figure \ref{hubble} plots the
average pitch angle \pave\ against the logarithm of the bulge-to-disk
ratio for each of these models. (Except for the model with $M_b=0$ of
course, which had pitch angles of $16.6^\circ$ and $23.1^\circ$ for
the halo rotation rates 0.075 and 0.105 respectively.)  The plot shows
a linear increase in \pave\ with $\log(M_b/M_d)$. As the bulge mass
increases, the spiral arms become tighter. The pitch angles for the
simulations with \omh=0.105 are systematically higher than those with
\omh=0.075. Plotted as well are data taken from Kennicutt (1981) for
the pitch angles of galaxies along the Hubble sequence, from Sa-type
to Sc-type. The bulge-to-disk ratios corresponding to these
classifications were taken from K\"oppen \& Arimoto (1990). The
scatter in the observations, mentioned at the top of this section, is
easily apparent.

The simulation data, however, follow the average trend of pitch angle
with Hubble type. The slopes of the simulation data are flatter than
that of the observations. Several explanations are viable for
explaining this: The simulated Hubble sequence created here is
artificial since it only varies one parameter, the bulge mass. There
could be systematic variations of total galaxy mass and length scales
along the Hubble sequence which could effect the spiral response of
the disk. Only two values of \omh\ were used, and the steeper slope of
the observations could be reproduced if it were assumed that higher
bulge mass correlated with lower halo rotation rates. Lastly, our
definition of \aave, since it is an average over a range of $\alpha$,
would tend to return values that are closer to the center of that
range, reducing the slope of the correlation. The pitch angles in
Kennicutt (1981) were measured by obtaining the arm coordinates from
H$\alpha$ or blue continuum observations and plotting these
coordinates on the log $r$ vs $\phi$ plane. A straight line was fitted
to each arm and the results for the two main arms were averaged.

\begin{figure}[t]
\vspace{11.5cm}
\includegraphics{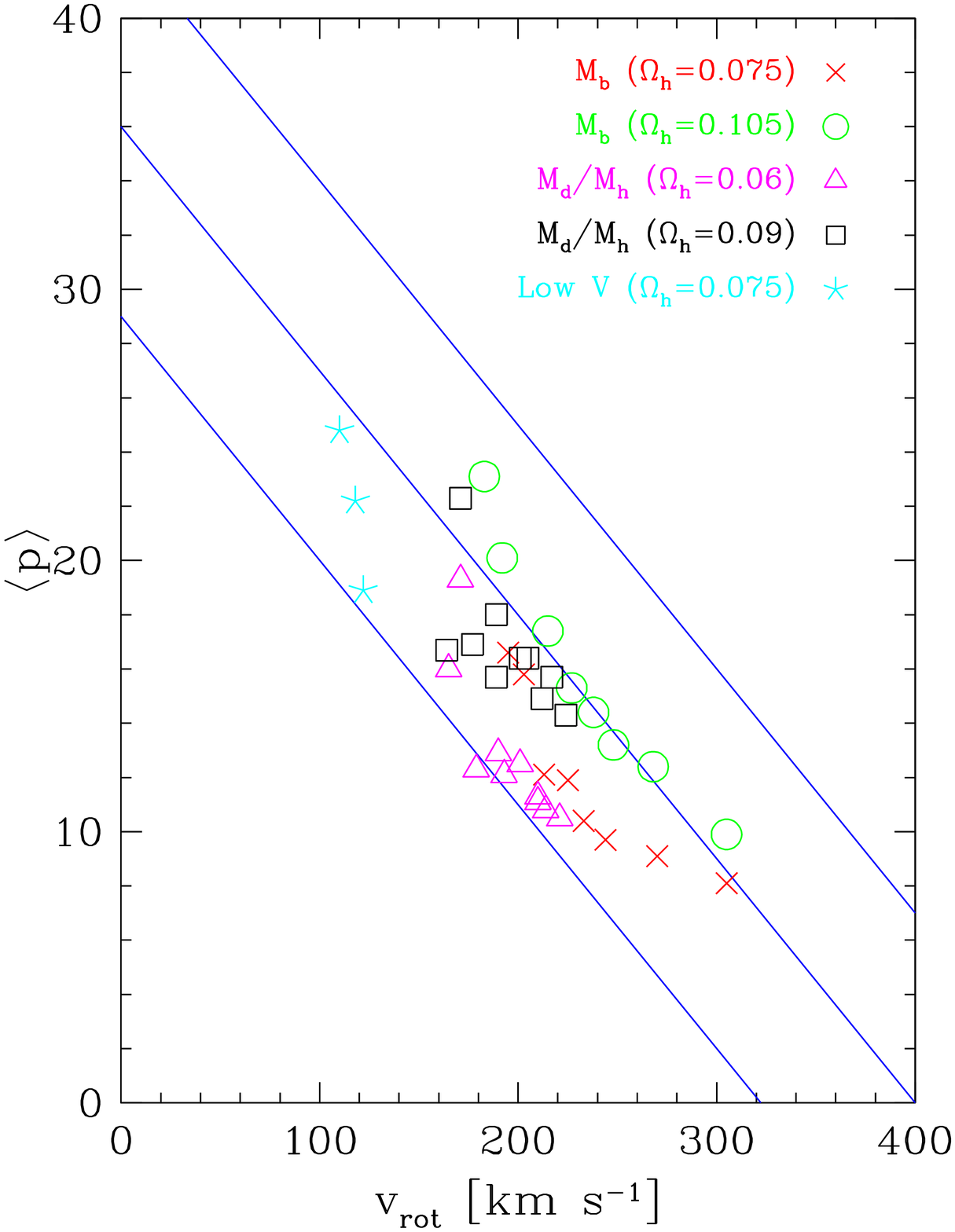}
\caption{\label{vrot} The average pitch angle, \pave, is plotted
against the maximum rotational velocity, \vrot, for each
simulation. Five sets of models are plotted: $M_b$ means a set where
the bulge mass is the varying quantity. $M_d/M_h$ means a set where
the mass ratio is varying. {\it Low V} means models designed to have
low rotational velocity. Descriptions of all these models can be found
in \S 4. The solid lines were taken from visual inspection of data in
Kennicutt (1981), representing the mean and the upper and lower bounds
of the majority of the observational data (excluding obvious
outliers).}
\end{figure}

\subsection{Pitch Angle and Rotational Velocity}

As noted above, a better predictor than Hubble classification of
spiral pitch angle is the rotational velocity of the galaxy (Kennicutt
1981). The relation found in the Kennicutt paper is plotted by the
solid lines in Figure \ref{vrot}. These lines represent the mean and
upper and lower bounds on his data. The slope and intercept of these
lines were calculated by inspection, and are not meant to be
quantitatively compared to our simulation results. Qualitatively,
however, it is obvious that our simulations follow the same
relation. Both sets of bulge simulations lie well within the bounds
set by the observations. What is remarkable as well is that the
simulations listed in Table 2 also show this trend. The simulations
with the lower rotation velocity generally have higher pitch
angles. Although the halo-dominated models do not return the same
pitch angles as the disk-dominated models with the same rotation
velocity, the trend is still preserved regardless of whether the mass
dominating the kinematics is from the halo or the disk or the
bulge. The scatter in our results lies within the scatter seen
observationally.

The lowest rotation velocity in the models listed in Table 2 is D5,
with $v_{rot}=165$ \kms. To explore the response of systems with even
lower velocities, we used the models described at the end of \S3 with
both low mass disks and halos. Three different bulge masses were used:
$M_b=$ 0, $0.15M_d$ and $0.3M_d$. The first two of these models developed a
strong and a weak bar respectively. The method of bar detection
used in \S4 allowed us to eliminate the particles that contribute to
the bar from being used in calculating \aave. Since the length of the
bar is known, all particles with radii below the bar length were
simply not counted in the summation in equation (12). The results are
plotted in Figure \ref{vrot}, and they follow the observational
relation.

\begin{figure}[t]
\vspace{10.5cm}
\includegraphics{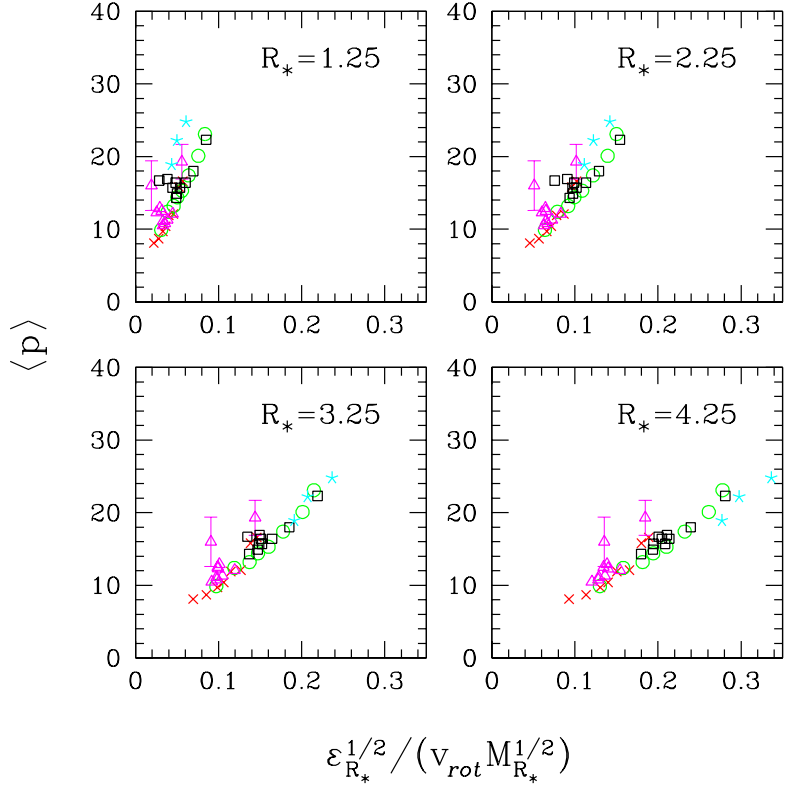}
\caption{\label{energy} The average pitch angle is plotted against the
normalized rotational energy of the halo for the same models shown in
Figure \ref{vrot}. The plot symbols are the same as Figure
\ref{vrot}. The radius $R_\ast$ at which \ep\ is calculated is shown
in the top right of each panel.  For $R_\ast=3.25$, all the model sets
form the tightest correlation, with the main outliers being from the
set with varying mass ratio with \omh\ = 0.06, which has the largest
error bars of all the sets. The points with error bars are the same
models in each panel. The error bars for the two outliers are larger
than the typical error by a factor of 2-3. The errors for \pave\ were
calculated by differential error on the relation between pitch angle
and $\alpha$, using the statistical error of \aavet. }
\end{figure}

A tighter correlation can be acheived by including what we know of the
mass distribution and halo rotation. Figure \ref{energy} plots \pave\
against the ``normalized'' rotational energy of the halo, a dimensionless
quantity given by

\begin{equation}
{\mathcal E}^\prime = {\mathcal E}_{R_\ast}^{1/2}/(v_{rot}M^{1/2}_{R_\ast}),
\end{equation}

\noindent where $\mathcal{E}$ is energy, $v_{rot}$ is the maximum
rotation velocity of the disk also used in Figure \ref{vrot}, and $M$
is the total mass of the galaxy. The energy in the right hand side of
equation (16) is the rotational energy of the halo, ${\mathcal
E}=I\Omega_h^2$, where $I$ is the moment of inertia of the halo. The
subscript $R_\ast$ on the energy and the total mass represents the
radius at which these two quantities are calculated. The four windows
in Figure \ref{energy} plot \pave\ against \ep\ calculated at four
different values of $R_\ast$. The correlation is apparent in all four
panels, but is excellent for $R_\ast=3.25$.

The choice of $R_\ast$ significantly influences the slope of the
correlation, with smaller values of $R_\ast$ resulting in steeper
slopes. The scatter changes only moderately with varying $R_\ast$ for
the first four sets of models in Figure \ref{energy}, but effects the
low-velocity models most. The quantity \ep\ has a power-law dependence
on $R_\ast$, which is highest with the low-velocity models. Thus the
values of \ep\ for that set will be affected most by changes in
$R_\ast$.

\section{Discussion}

We have presented simulations of galaxy disks embedded in rigid,
rotating, triaxial halos. We have constructed a suite of models in
which the masses of the disk, bulge and halo have all been
varied. With these models, the effects of changing the halo rotation
rate and flattening have been investigated. Our results have focused
on the formation of bars and spiral structure. Models with a mass
ratio \mratio\ $> 0.8$ and \omh\ $=0.09$ develop a strong bar, even if
a significant bulge is present. For all models, the torque created by
the halo induces spiral structure in the disk. The average pitch angle
of this structure, \pave, is strongly anti-correlated with the maximum
rotation speed of the model.

It is clear that estimations of the stability of disk galaxies cannot
ignore the shape of the halo. Models with significant amounts of dark
matter and a significant bulge (like D1 and D2, which have rotational
velocities of over 200 \kms\ in our system of units) can become bar
dominated with halo rotation rates of \omh\ = 0.09 and
greater. Maximal disk models, like D5, are very susceptible to bar
instabilities if torqued by the halo. In all the simulations presented
in \S 5, the only set of halo and bulge parameters for which models
D3--D5 were stable against bar formation was $M_b=0.3M_d$ and \omh\ =
0.06. Even with those parameters, model D3 still developed a low
amplitude but persistent bar. A simple criterion for determining
whether a disk galaxy will develop a bar, such as that determined by
Efstathiou \etal\ (1982), is not readily apparent from the data. With
more free parameters, $M_b$, $q_b$, and \omh, in our simulations, it
may be that a simple criterion cannot quantify the stability of all
our models used in our simulations.

The correlation of rotational velocity and pitch angle in our models
provides a physical explanation of the same correlation found in
observations of spiral galaxies. The pitch angle of the arms induced
by an external torque is related to the total mass of the system,
loosely through $v_{max}$, but more tightly to the dimensionless
rotational energy parameter \ep. The differences seen in our results for
different halo rotation rates and flattenings fit well with the amount
of scatter seen in the correlation of $v_{max}$ with \pave\ for real
galaxies. This result does not have to be unique to the properties of
a galaxy's halo; it has been shown that encounters with satellite
galaxies, which would also create a time-varying potential field in
which to disrupt the disk, induce two-armed spiral structure in disk
galaxies (\eg\ Toomre \& Toomre 1972). But encounters with smaller
galaxies are by their nature transient and it remains that spiral
structure is a ubiquitous feature of disk galaxies both with and
without satellite perturbers.

In Kennicutt (1981), and references therein, the properties of the
spiral arms are measured through their blue light, usually from
H{\small II} regions, dust lanes, or the blue continuum, all of which
are associated with star formation. Our collisionless simulations
better represent the red stellar population which accounts for most of
the mass in a galactic disk. Images of spiral galaxies in the near
infrared have shown that spiral structure is evident in old stellar
populations (Eskridge \etal\ 2002). The arms observed are smoother and
wider than those seen in blue light, but generally follow the same
morphology. 

The Fourier decomposition method used in this paper has been applied
to observational data of H{\small II} regions in spiral galaxies
(Puerari \& Dottori 1992; Garc\'ia-Gom\'ez \& Athanssoula 1993). The
Fourier spectra obtained resemble in many ways the spectra of our
simulated galaxies, often with several distinct peaks inside one
spectrum. The pitch angle was calculated by these authors from the
value of $\alpha$ for which the amplitude is highest. This definition
of $p$ reproduces the general trend with Hubble type as described by
Kennicutt (1981) but does not reflect the natural variations in pitch
angle within a single galaxy seen both in the Fourier spectra of the
observations and in Kennicutt's calculations. Kennicutt noted that the
variations of $p$ for a typical galaxy were significantly larger than
the measurement error. An average value of the pitch angle better
represents the multiple spiral components.

Hydrodynamic simulations by Bekki \& Freeman (2002) have shown that
the extended spiral structure seen in the blue compact dwarf galaxy
NGC 2915 could be due to a rotating triaxial halo. Their simulations
contained a disk fully composed of gas particles and a halo 100 times
more massive than the disk, effectively making the disk massless. Our
collisionless simulations show spiral structure in the inner regions
of galactic disks. Since gas fractions in disk galaxies are usually no
higher than 30\%, the gas should be gravitationally coupled to the
stars and follow the spiral patterns seen in our simulations. The
dissipation and subsequent star formation in the gas would lead to
self-consistent spiral structure which would outlast a collisionless
disk and work to preserve its structure as well.

An open question is the correct value or range of values to use for
\omh. The figure rotation of NGC 2915 has been calculated to be 8.0
\kmskpc\ (Bureau \etal\ 1999), which is 0.107 in our $N$-body
units. These authors also propose that halo figure rotation, a
distinct phenomenon from halo angular momentum, is present in a
significant fraction of dark matter halos in CDM simulations.

Regardless of the the details in determining \omh\ or \pave, it has
been clearly demonstrated that an external torqueing mechanism can
reproduce the type of correlation observed between spiral pitch angle
and rotational velocity. Even in simulations in which the magnitude of
the applied torque is the same, \ie\ the density profile, flattening,
and rotation rate of that halo are all held constant, the pitch angle
of the spiral structure varies according to the total mass of the
system.

\acknowledgements JLT would like to thank Richard Pogge for his
helpful discussions, as well as Volker Springel and Stein Sigurdsson
for help with the \gadget\ and \scf\ codes respectively. This work was
performed at the Ohio Supercomputer Center under grant PAS0825.

\clearpage


\begin{references}

Alam, S.M.K. \& Ryden, B.S. 2002, ApJ, 576, 610

Barnes, J.E. \& Hut, P. 1986, Nature, 324, 446

Barnes, J.E. 1992, ApJ, 393, 484

Binney, J. \& Tremaine, S. 1987 Galactic Dynamics (Princeton
Univ. Press)

Bureau, M., Freeman, K.C., Pfitzner, D.W., \& Meurer, G.R. 1999, AJ,
118, 2158

Dubinski, J. \& Carlberg, R.G. 1991, ApJ, 378, 996

Dubinski, J. 1994 ApJ, 431, 617

Dubinski, J., Mihos, J.C., \& Hernquist, L. 1999, ApJ, 526, 607

Efstathiou, G., Lake, G., \& Negroponte, J. 1982, MNRAS, 199, 1069

Eskridge, P.B., \etal\ 2002, ApJS, accepted, (astro-ph/0206320)

Garc\'ia-Gom\'ez,  C. \& Athanassoula, E. 1993, A\&AS, 100, 431

Holh, F. 1976, AJ, 81, 30

Hernquist, L. 1990, ApJ, 356, 359

Hernquist, L. 1993, ApJS, 86, 389

Hernquist, L. \& Ostriker, J.P. 1992, ApJ, 386, 375

Holley-Bockelmann, K., Mihos, J.C., Sigurdsson, S., \& Hernquist,
L. 2001, ApJ, 549, 862

Huntley, J.M., Sanders, R.H., \& Roberts, W.W. 1978, ApJ, 221, 521

Ibata, R., Lewis, G.F., Irwin, M., Totten, E., \& Quinn, T. 2001, ApJ,
551, 294

Jing, Y.P. \& Suto, Y. 2002, ApJ, 574, 538

Kennicutt, R.C. 1981, AJ, 86, 1847

Kennicutt, R.C. \& Hodge, P. 1982, ApJ, 253, 101

K\"oppen, J. \& Arimoto, N. 1990, A\&A, 240, 22

Lin, C.C. 1970, in IAU Symposium No. 38: The Spiral Structure of our
Galaxy, ed. Becker and Contopoulos (Dordrecht: Reidel), p 377

Mihos, J.C. \& Hernquist, L. 1996, ApJ, 464, 641

Merritt, D. \& Quinlan G.D. 1998, ApJ, 498, 625

Murali, C., Katz, N., Hernquist, L., Weinberg, D.H., \& Dav\'e,
R. 2002, ApJ, 571, 1

Navarro, J.F., Frenk, C.S., \& White, S.D.M. 1997, ApJ, 490, 493

Ostriker, J.P. \& Peebles, P.J.E. 1973, ApJ, 186, 467

Puerari, I. \& Dottori, H.A. 1992, A\&AS, 93, 469

Ryden, B.S. 1992, ApJ, 396, 445

Ryden, B.S. 1996, ApJ, 461, 146

Sanders, R.H. \& Huntley, J.M. 1976, ApJ, 209, 53

Sanders, R.H. 1977, ApJ, 277, 916

Sanders, R.H. \& Tubbs A.D. 1980, ApJ, 235, 803

Sellwood, J.A. \& Athanassoula, E. 1986, MNRAS, 221, 195

Sellwood, J.A. \& Sparke, L.S. 1988, 231, 25

Siegel, M.H., Majewski, S.R., Reid, I.N., \& Thompson, I.B. 2002, ApJ,
accepted (astro-ph/0206323)

Springel, V. 2000, MNRAS, 312, 859

Springel, V., Yoshida, N., \& White, S.D.M. 2001, NewA, 6, 79

Thakar, A.R. \& Ryden, B.S. 1998, ApJ, 506, 93

Toomre, A. \& Toomre, J. 1972, ApJ, 178, 623

Tremblay, B. \& Merritt, D. 1996, AJ, 111, 2243

Warren, M.S., Quinn, P.J., Salmon, J.K., \& Zurek, W.H. 1992, ApJ,
399, 405

\end{references}
\end{document}